\documentclass[aps,twocolumn,showpacs,superscriptaddress,footnoteinbib, prl, reprint]{revtex4-2}
\usepackage{amsmath, amsfonts, amssymb, wasysym, mathrsfs, dsfont}
\usepackage{graphicx, color}
\usepackage{ulem} 
\usepackage{dcolumn} 
\usepackage[dvipsnames]{xcolor}
\usepackage{xspace}
\usepackage{nicefrac, bm}
\usepackage[colorlinks, breaklinks, linkcolor=OrangeRed, citecolor=RoyalBlue, urlcolor=NavyBlue]{hyperref}
\usepackage[all]{hypcap} 

\newcommand{\R}{\ensuremath{\mathbb{R}}}
\newcommand{\Z}{\ensuremath{\mathbb{Z}}}

\newcommand{\di}[1]{\ensuremath{\mathrm{d}#1 \,}}

\newcommand{\1}{\ensuremath{\mathbb{1}}}

\newcommand{\trsp}{\ensuremath{{\vbox{\hbox{$\scriptstyle\intercal$}\nointerlineskip\hbox{}}}}}
\newcommand{\cre}[1]{\hat{a}^\dagger_{#1}}
\newcommand{\anh}[1]{\hat{a}^{\phantom{\dagger}}_{#1}}
\newcommand{\creb}[1]{\hat{b}^\dagger_{#1}}
\newcommand{\anhb}[1]{\hat{b}^{\phantom{\dagger}}_{#1}}

\newcommand{\ddroit}[2]{\ensuremath{\dfrac{\mathrm{d}{#1}}{\mathrm{d}{#2}}}}

\newcommand{\ket}[1]{\ensuremath{|#1\rangle}\xspace}
\newcommand{\bra}[1]{\ensuremath{\langle #1|}\xspace}

\newcommand{\bracket}[2]{\ensuremath{\langle {#1} | {#2} \rangle}}
\newcommand{\avrg}[1]{\ensuremath{\left \langle #1 \right \rangle}}

\newcommand{\vertiii}[1]{{\left\vert\kern-0.25ex\left\vert\kern-0.25ex\left\vert #1
    \right\vert\kern-0.25ex\right\vert\kern-0.25ex\right\vert}} 

\DeclareFontFamily{U}{wncy}{}
\DeclareFontShape{U}{wncy}{m}{n}{<->wncyr10}{}
\DeclareSymbolFont{mcy}{U}{wncy}{m}{n}
\DeclareMathSymbol{\sha}{\mathord}{mcy}{"58}

\begin{document}

\title{Spin-boson quantum phase transition in multilevel superconducting qubits}


\author{Kuljeet Kaur}
\affiliation{Department of Physics, Indian Institute of Technology Bombay, Mumbai 400076, India}
\author{Th\'eo S\'epulcre}
\affiliation{Univ. Grenoble Alpes, CNRS, Institut N\'eel, F-38000 Grenoble, France}
\author{Nicolas Roch}
\affiliation{Univ. Grenoble Alpes, CNRS, Institut N\'eel, F-38000 Grenoble, France}
\author{Izak Snyman}
\affiliation{Mandelstam Institute for Theoretical Physics, School of Physics, University of the Witwatersrand,
Johannesburg, South Africa}
\author{Serge Florens}
\affiliation{Univ. Grenoble Alpes, CNRS, Institut N\'eel, F-38000 Grenoble, France}
\author{Soumya Bera}
\affiliation{Department of Physics, Indian Institute of Technology Bombay, Mumbai 400076, India}

\begin{abstract}
Superconducting circuits are currently developed as a versatile platform for the exploration of
many-body physics, by building on non-linear elements that are often idealized as two-level qubits.
A classic example is given by a charge qubit that is capacitively coupled to a transmission line, 
which leads to the celebrated spin-boson description of quantum dissipation. We show that the intrinsic 
multilevel structure of superconducting qubits drastically restricts the validity of the spin-boson 
paradigm due to phase localization, which spreads the wavefunction over many
charge states.
Numerical Renormalization Group simulations also show that the quantum critical point moves out of the 
physically accessible range in the multilevel regime. Imposing charge discreteness in a simple variational 
state accounts for these multilevel effects, that are relevant for a large class of devices.

\end{abstract}
\maketitle

Quantum computation has been hailed as a promising avenue to tackle a large
class of unsolved problems, from physics and
chemistry~\cite{QuantumComputationalChemistry} to algorithmic
complexity~\cite{QuantumAlgorithms}. This research follows an original proposition from
Feynman~\cite{FeynmanQuantum},
long before the technological and conceptual tools were developed to make such ideas
tangible~\cite{ReviewQuantumComputing}.
While a general purpose digital quantum computer could theoretically
outperform classical hardware for some exponentially hard tasks, building such a
complex quantum machine is at present out of reach. For this reason,
analog quantum simulation has been put forward as a crucial milestone~\cite{Nori_RMP},
aiming at the design of fully controllable experimental devices mimicking the
features of difficult quantum problems of interest. This route has met tremendous success
in the past, with the realization of Kondo impurities in quantum dots~\cite{Kouwenhoven:2001wd},
the simulation of artificial solids in optical lattices~\cite{Bloch:2008gl}, and is gaining
momentum with new tools from superconducting
circuits~\cite{Houck:2012iq, Raftery:2014jk, FornDiaz:2016bo, Roushan:2017ij,
Ma:2019aa, Leger:2019vv, Carusotto:2020ue}.
Ironically, while Feynman anticipated quantum simulators~\cite{FeynmanQuantum},
he often warned in his lectures (where analogy was used as a powerful teaching method)
that there is no such thing as a perfect analogue, and that some interesting physics
can emerge when the analogy breaks down~\cite{FeynmanLecture}.
Exploring realistic superconducting circuits for the emulation of strongly interacting 
quantum spin systems is the main purpose of this Letter. By underlining the crucial role of 
multilevel effects, we aim to unveil the peculiar many-body physics of such simulators. Our
study will focus on the realm of quantum dissipation~\cite{Leggett_RMP,Weiss}, a problem that 
is still raising increasing interest~ \cite{FornDiaz:2016bo, PuertasMartinez:2019gk, kuzmin_superstrong_2019} 
due to potential applications ranging from hardware-protected qubits \cite{Weiss:2019io, DiPaolo:2019wl} 
to quantum optics with metamaterials~\cite{Grimsmo:2020va}. Many ideas that will be presented here will 
however apply to the more general context of superconducting simulators of many-body
problems~\cite{ReviewSuperconductingSimulators}.

Addressing the full complexity of superconducting circuit simulators raises a long list of theoretical 
challenges, and we emphasize already now the four unsolved issues related to multilevel physics that we tackle in this Letter.
{\bf (1)} Most quantum simulation protocols assume that qubits behave as idealized spin 1/2 degrees of freedom. 
While a large class of mesoscopic systems fall under this 
assumption~\cite{Cedraschi_Ring,Matveev_Resonant,LeHur_Resonant,LeHur_Box,LeHur_Caldeira,Recati_Atomic},
this is clearly questionable for superconducting qubits where the non-linearity is only provided by the
cosine Josephson potential. Indeed, we show that the two-level description can be invalid 
for many-body ground states due to proliferation of multilevel states (at strong driving~\cite{Verney}, 
multilevel effects are known to even plague a single Josephson junction).
{\bf (2)} Quantitative modeling of simulators involving a large number of qubits or resonators
requires to incorporate the full capacitance network of the circuit
\cite{Anonymous:2012ek,GarciaRipoll:2015ba, malekakhlagh_cutoff-free_2017, ParraRodriguez:2018da,
PuertasMartinez:2019gk}. We will see that such electrokinetic considerations impose strong constraints
for models based on multilevel qubits, that can even prevent the occurence of quantum phase transitions.
{\bf (3)} Effects beyond the simple RWA approximation can be difficult to simulate numerically 
due to the exponential size of the Hilbert space, making the study of many-body dissipation
challenging~\cite{FornDiaz:2019br,Kockum:2019ky,LeBoite}. We find however that handling the complete 
multilevel structure of Josephson qubits can be tackled by the Numerical Renormalization Group (NRG).
{\bf (4)} Finally, random charge offsets are a notorious experimental nuisance for the operation of 
superconducting circuits in strongly non-linear regimes, but are also difficult to model, because they cannot 
be captured by a Kerr expansion~\cite{Koch_Transmon,Verney}. We propose here a simple wavefunction encoding multilevel 
charge discreteness, that remarkably reproduces our full NRG simulations. This simple analytical 
theory can be extended beyond dissipative models, {\it e.g.} to study bulk quantum
phase transitions~\cite{GlazmanLarkin,Basko_SIT,Roy_SineGordonSimulation} in presence charge noise~\cite{InPrep}.

\begin{figure}[ht]
    \centering
    \includegraphics[width=1.0\columnwidth]{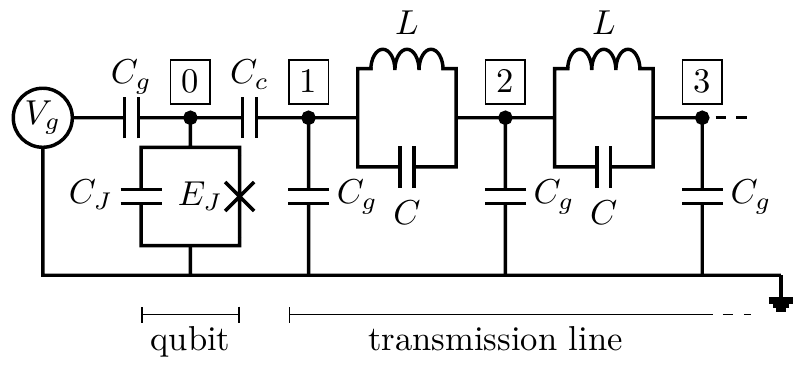}
    \caption{Microscopic electrokinetic model for a 
realistic circuit of
dissipative superconducting charge qubit,
located at node 0, characterized by Josephson energy $E_J$,
shunt capacitance $C_s$, and capacitively coupled via $C_c$ to a
transmission line. All nodes are shunted to the ground via the capacitance $C_g$, and each
lumped element in the line is characterized by its inductance $L$ and self-capacitance $C$.
Charge offsets are modeled by a DC voltage source $V_g$.}
    \label{fig:circ}
\end{figure}

Numerous theoretical works have recently studied the ultra-strong coupling physics of superconducting qubits, 
based on the two-level approximation~\cite{LeHur_Kondo_2012,goldstein_inelastic_2013,
peropadre_2013,snyman_robust_2015,Gheeraert:2018bv,Magazzu}. While this assumption is valid for the Cooper 
pair box (a qubit designed with strong charging energy), this regime is however unfavorable experimentally 
due to high sensitivity to external charge noise.
A more realistic circuit is shown in Fig.~\ref{fig:circ}, composed 
of a superconducting charge qubit containing a junction with Josephson energy $E_J$ and 
capacitance $C_J=C_{s}+C_{g}$ where $C_{s}$ is a shunt capacitance and $C_{g}$ is a gate capacitance,
that is capacitively coupled via $C_c$ to a transmission line characterized by lumped element
inductance $L$ and capacitance $C$. All nodes are grounded \textit{via} capacitances $C_g$, and
a DC charge offset controlled by voltage $V_g$  is included on qubit node $0$,
appearing as dimensionless charge $n_g=V_gC_{g}/2e$.
We will not make any assumption on all these parameters here.
The transmission line may be designed in practice from an array of linear Josephson 
elements~\cite{PuertasMartinez:2019gk}, in order to boost its coupling to the qubit, thanks to the
high optical index $n\simeq100$ which slows down accordingly the velocity of microwave modes.
The circuit Lagrangian reads~\cite{vool_introduction_2017} (working in units
of $\hbar=2e=1$):
\begin{equation}
    \mathcal{L} = \frac{1}{2}\dot{\vec{\Phi}}^{\, \trsp} \bm{C}\dot{\vec{\Phi}}
- \frac{1}{2}\vec{\Phi}^{\, \trsp} \nicefrac{\bm 1}{\bm L}\vec{\Phi} + E_J
  \cos(\Phi_0)
- n_g\dot{\Phi}_0, \label{eq:startL}
\end{equation}
where $\vec{\Phi} = (\Phi_0, \Phi_1, \hdots)$ is a vector of
dimensionless node fluxes labeled according to Fig.~\ref{fig:circ}. $\bm C$ and
$\nicefrac{\bm 1}{\bm L}$ are the capacitance and inductance matrices read from
Fig.~\ref{fig:circ}, that
define a generalized eigenvalue problem
$\nicefrac{\bm 1}{\bm L}\bm{P} = \bm{CP\omega}^2$, $\bm \omega$ being the
diagonal matrix of the system eigenfrequencies, bringing the Lagrangian in
normal mode form in the new basis $\vec{\phi} = \bm{P}^{-1}\vec{\Phi}$. 
The qubit degree of freedom can be separated from the external modes via the change of variables
$ \varphi = \sum_k P_{0k}\phi_k$ and $\varphi_m = \phi_m$~\cite{SupInfo}.
Once the bath modes $\varphi_m$ are quantized in terms of creation/annihilation operators, we obtain 
the Hamiltonian:
\begin{eqnarray}
  \hat{H} &=& \sum_k \omega_k \cre{k}\anh{k} +\, (\hat{n} - n_g)\sum_k ig_k(\cre{k} - \anh{k}) \nonumber \\
\vspace{-0.3cm}
  & &
+\, 4 E_c (\hat{n}-n_g)^2 - E_J\cos\hat{\varphi}  , \label{eq:fullH}
\end{eqnarray}
which we name the ``charge-boson model'', as the charging 
$4 E_c (\hat{n}-n_g)^2 = 4 E_c \sum_n(\big|n\big>\big<n\big|-n_g)^2$ and Josephson
energy $E_J \cos\hat{\varphi} = (E_J/2) \sum_n(\big|n\big>\big<n+1\big|+\mathrm{h.c.})$ 
are represented in the full multilevel charge basis $\{\big|n\big>\}$ with $n\in\mathbb{Z}$.
This model generalizes to many levels the standard two-level ``spin-boson model'' describing 
quantum dissipation~\cite{Leggett_RMP,Weiss}. Indeed, for a Cooper pair box in the regime $E_c/E_J\gg1$, one can truncate 
the full spectrum of the Josephson junction to the two charge states closest to $n_g$, namely 
$n_0=\left \lfloor{n_g}\right \rfloor$ and $n_0+1$, 
so that the charge operator reads
$\hat{n} \simeq 
n_0\big|\!\!\uparrow\!\!\big>\big<\!\!\uparrow\!\!\big| +
(n_0+1)|\!\!\downarrow\!\!\big>\big<\!\!\downarrow\!\!|$, while
$\cos \hat{\varphi} \simeq (\big|\!\!\uparrow\!\!\big>\big
<\!\!\downarrow\!\!| + |\!\!\downarrow\!\!\big>\big<\!\!\uparrow\!\!|)/2 = \hat\sigma_x/2$.
At the charge degeneracy point $n_g=1/2$+integer, we find 
$\hat{n}-n_g\simeq \hat\sigma_z/2$, so that 
(\ref{eq:fullH}) takes the usual spin-boson form.

In the charge-boson Hamiltonian~(\ref{eq:fullH}), $g_k = \sqrt{\omega_k/2}P_{0k}$ are the couplings to 
the bosonic normal modes, and the qubit charging energy obeys:
\begin{equation}
4E_c = \frac{P_{00}^2}{2} + \sum_{k\neq0} \frac{g_k^2}{\omega_k}
= \frac{1}{2C_J+2C_c} +
\frac{1}{\pi}\int_0^\infty\!\!\!\!\! \mathrm{d}\omega\; \frac{J(\omega)}{\omega}.
\label{eq:charging}
\end{equation}
Here $J(\omega) = \pi \sum_k g_k^2 \delta(\omega_k -
\omega)$
lumps together the couplings to all modes into a spectral function
that is smooth for an infinite chain~\cite{Leggett_RMP,Weiss}.
For the circuit of Fig.~\ref{fig:circ}, $J(\omega) = 
2\pi\alpha \omega \sqrt{1\!-\!\omega^2/\omega_P^2}/(1\!+\!\omega^2/\omega_J^2)
\theta(\omega_P\!-\!\omega)$,
with dissipation strength $2\pi\alpha=(4e^2/h)[C_c/(C_c+C_J)]^2 \sqrt{L/C_g}$,
plasma frequency $\omega_P=1/\sqrt{L(C+C_g/4)}$,
and a non-trivial RC cutoff of the junction $\omega_J=1/\sqrt{LC_\mathrm{eff}}$ with 
$C_\mathrm{eff}=C_JC_c/(C_J+C_c)+(C_JC_c)^2/[C_g(C_J+C_c)^2]-C$~\cite{SupInfo}.

In order to unveil the crucial role of the multilevel structure in the dissipative 
Hamiltonian~(\ref{eq:fullH}), we eliminate the capacitive coupling via a unitary transform
$\hat{U} = \exp[i(\hat{n}-n_g)\sum_k (g_k/\omega_k)(\cre{k}+\anh{k})]$, resulting in:
\begin{eqnarray}
\hat{U}\hat{H}\hat{U}^\dagger &=& \sum_k \omega_k \cre{k}\anh{k} 
+\, \left( 4 E_c-
\frac{1}{\pi}\int_0^\infty\!\!\!\!\! \mathrm{d}\omega\; \frac{J(\omega)}{\omega}
\right) (\hat{n}-n_g)^2 \nonumber \\
\vspace{-0.3cm}
  & &
- E_J\cos\Big[\hat{\varphi}- \sum_k (g_k/\omega_k) (\cre{k} + \anh{k})\Big] . \label{eq:newH}
\end{eqnarray}
This expression shows that the charging energy $E_c$ and the spectral function $J(\omega)$ of
the environment are not independent parameters, since taking $E_c$ down to zero would result
in a negative capacitance. Indeed, Eq.~(\ref{eq:charging}) clearly shows that the 
capacitance always stays positive. However, the constraint $4 E_c > 
\frac{1}{\pi}\int_0^\infty\! \mathrm{d}\omega J(\omega)/\omega$
becomes hidden upon making the two-level approximation, since the quadratic charging 
term disappears in the spin-boson model when taking the limit $E_c\to\infty$.
This implies that the dissipation strength $\alpha$ has an upper bound:
\begin{equation}
    \alpha \leqslant \alpha_{\mathrm{max}}=2E_c/\omega_c. \label{eq:alphaMax}
\end{equation}
with $\omega_c \simeq \mathrm{Min}(\omega_P,|\omega_J|)$, as obtained
by parametrizing the spectral function as
$J(\omega)=2 \pi \alpha \omega \exp{(-{\omega}/{\omega_c})}$.
Such electrostatic constraint must be fulfilled for any microscopic model, and 
we provide the exact bound for the circuit of Fig.~\ref{fig:circ} in~\cite{SupInfo}.
From Eq.~(\ref{eq:charging}), the maximum value of dissipation $\alpha_\mathrm{max}$ is 
attained for $C_c\to\infty$, namely when the qubit becomes wire coupled to the transmission 
line (see Fig.~\ref{fig:circ}). In that case, charge quantization is lost and the transformed 
charge-boson Hamiltonian~(\ref{eq:newH}) becomes equivalent to the boundary sine-Gordon 
model~\cite{Zwerger_Periodicity}, because the phase $\hat{\varphi}$ obviously freezes 
out, leaving the cosine potential as a boundary effect on the bosonic modes.
We emphasize that the resulting Schmid transition~\cite{SchoenZaikin} has a different 
universality to the spin-boson transition that we study, and is not relevant for the case 
of finite $C_c$ considered here.

The constraint~(\ref{eq:alphaMax}) has
profound consequences for the dissipative quantum mechanics of realistic
charge qubits. Indeed, reaching the ultrastrong coupling regime
$\alpha \simeq1$ where many-body effects are most prominent implies
$E_c\simeq \omega_c$. For Cooper pair boxes with $E_J\ll E_c$, the 
first excited qubit state lies at energy $E_J\ll\omega_c$, well within the linear
regime of $J(\omega)$, so that ohmic dissipation controls the qubit dynamics, allowing
the ohmic spin-boson transition~\cite{Leggett_RMP}.  
In the other extreme regime $E_J\gg E_c$, the first qubit excitation
located at $\sqrt{8E_c E_J}\gg\omega_c$ now lies in the tails of the cutoff function $J(\omega)$. 
This suggest that the ohmic spin-boson quantum phase transition is not possible 
for a capacitively coupled transmon qubit, shedding light on previous experimental attempts
\cite{PuertasMartinez:2019gk, kuzmin_superstrong_2019}, and extending predictions for 
systems of transmons coupled to single cavities~\cite{Bosman:2017el,Jaako:2016io,DickeNoGo}.
Establishing at which value of $E_J/E_c$ the phase transition becomes
forbidden in the full charge-boson model is very important to guide experimental
endeavors on superconducting simulators, and requires a full-fledged many-body
solution of the problem.
For this purpose, we first need to uncover the order parameter controlling the
quantum phase transition in the charge-boson model. Due to the periodicity of charge quantization, 
we can restrict $n_g\in [0,1]$.
We notice that, for $n_g=1/2$, Hamiltonian~(\ref{eq:fullH}) is
invariant by the symmetry:
$a^\dagger_k \to - a^\dagger_k$ and
$\hat{n} \to 1-\hat{n}$. 
If the ground state of Hamiltonian~(\ref{eq:fullH}) preserves this symmetry,
we get $\big<\hat{n}\big>=\big<1-\hat{n}\big>$, so that $\big<\hat{n}\big>=1/2$, namely
$\big<\hat{n}-n_g\big>=0$.
On the contrary, if the symmetry is spontaneously broken, $\big<\hat{n}-n_g\big>\neq0$ 
serves as an order parameter. This is physically expected
because the linear coupling term $(\hat{n} - n_g)\sum_k ig_k(\cre{k} - \anh{k})$ 
in Eq.(\ref{eq:fullH}) tends to induce a finite
charge polarization $\hat{n}-n_g$.

\begin{figure}[htbp]
    \centering
    \includegraphics[width=1.0\columnwidth]{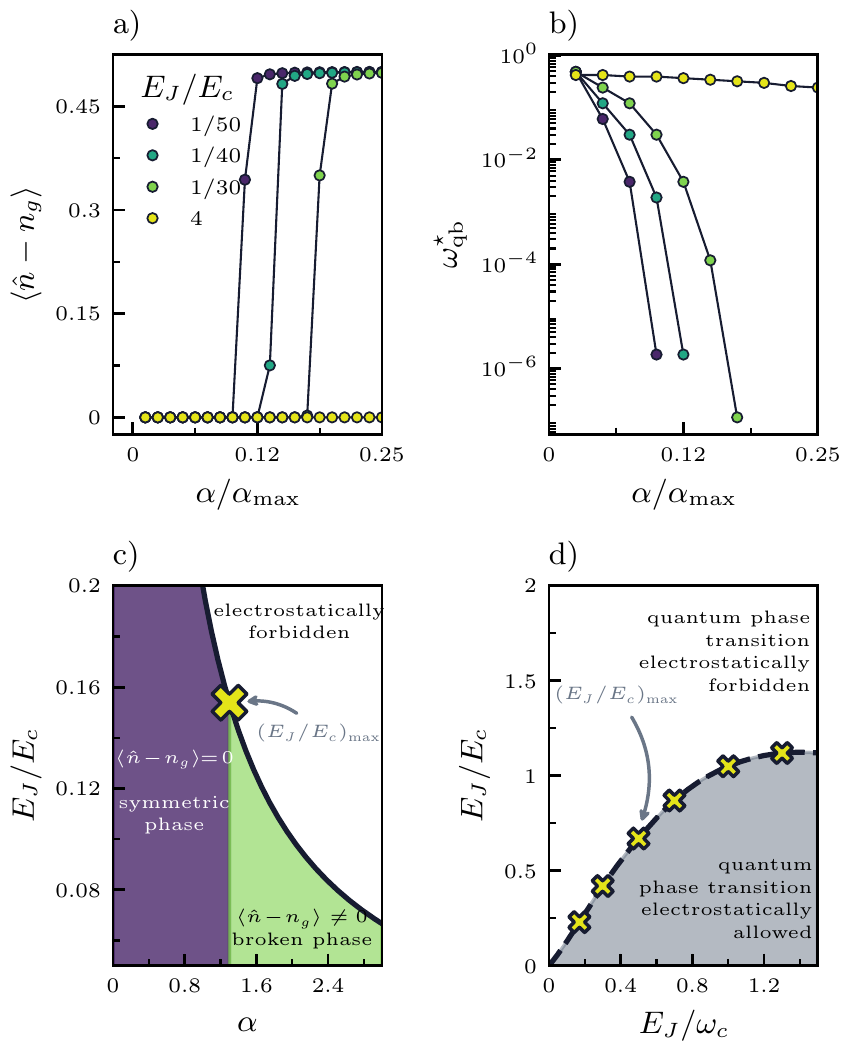}
\vspace{-0.5cm}
    \caption{{\bf a)} Order parameter $\big<\hat{n}-n_g\big>$ as a function of dissipation $\alpha$
(normalized to $\alpha_\mathrm{max}$) for four values of $E_J/E_c$, obtained at fixed $E_J/\omega_c=0.1$ 
and at the degeneracy point $n_g=1/2$.  A quantum phase transition is only obtained if $E_J/E_c\ll1$. 
{\bf b)} Renormalized qubit frequency $\omega_\mathrm{qb}^\star$ for the same parameters, which vanishes
at the same critical point. {\bf c)}
Phase diagram of the charge-boson model showing the phase boundary and the electrostatic
forbidden regime ($E_J/\omega_c=0.1$ is fixed).
{\bf d)} General phase diagram for arbitrary $E_J/E_c$ and $E_J/\omega_c$, showing regimes
where a spin-boson quantum phase transition for multilevel qubits is ruled out.}
\label{fig:QPTresult}
\end{figure}

Computing the order parameter $\big<\hat{n}-n_g\big>$ can only be achieved
by reliable quantum many-body simulations of the charge-boson Hamiltonian~\eqref{eq:fullH}. 
Taking advantage of the impurity structure of
the problem, we have extended the Numerical Renormalization Group (NRG)~\cite{NRG-RMP08} to
dissipative Josephson junctions, in contrast to previous treatments of the
spin-boson model based on the two-level system approximation~\cite{Tong}.
The method is based on an iterative diagonalization, adding modes one by one
on a logarithmic grid, with a truncation of the Hilbert space at each NRG step. 
For the charge-boson model~\eqref{eq:fullH}, the first stage of the NRG starts 
with the qubit degree of freedom, expressed in the charge basis, with up to 
$10^3$ charge states to ensure proper convergence for all considered $E_J/E_c$ values. 
We work with the Ohmic model $J(\omega) = 2\pi\alpha\omega
\exp{(-\omega/\omega_c)}$ in units of $\omega_c=1$, and start the NRG procedure
with frequencies of order $10\,\omega_c$ down to the minimal frequency
$10^{-14}\omega_c$ that guarantees convergence of the NRG to the full many-body
ground state.

Our first important finding concerns the dissipation-induced 
quantum phase transition of the charge-boson Hamiltonian~(\ref{eq:fullH}), beyond
the two-level approximation.
Fig.~\ref{fig:QPTresult}a shows the ground state order parameter
$\big< \hat n -n_g \big>$ as a function of normalized dissipation $\alpha/\alpha_\mathrm{max}$, 
which always stays zero when $E_J> E_c$. However,
Cooper pair box qubits with $E_c\gg E_J$ do show a transition.
This scenario is confirmed by monitoring the enhanced quantum fluctuations
in the symmetric phase, from the charge response function 
$\chi(t)=\big< (\hat{n}(t)-n_g)(\hat{n}(0)-n_g)\big>$ of
the qubit.  
A peak in the frequency domain occurs at the scale $\omega_\mathrm{qb}^\star$,
associated to the renormalized frequency of the qubit. Extracting $\omega_\mathrm{qb}^\star$
for various parameter values, we see in Fig.~\ref{fig:QPTresult}b that $\omega_\mathrm{qb}^\star$ 
vanishes exponentially fast at the quantum critial point.
Drawing the resulting phase diagram in the $(\alpha,E_J/E_c)$ plane (here $E_J/\omega_c=0.1$ is 
fixed), we find in Fig.~\ref{fig:QPTresult}c that the transition point between the two phases 
simply disappears when $E_J/E_c$ is increased
(cross), due to the border to the electrostatically forbidden region. Reporting the boundary 
$(E_J/E_c)_\mathrm{max}$ in the $(E_J/E_c, E_J/\omega_c)$ plane, we obtain a completely general 
phase diagram in Fig.~\ref{fig:QPTresult}d.
We thus established that the regime $E_J/E_c \gtrsim 1$ always forbids quantum criticality, so 
that the spin-boson paradigm does not apply for multilevel charge qubits, including transmons 
($E_J\gg E_c$).
\begin{figure}[htbp]
    \centering
    \includegraphics[width=0.99\columnwidth]{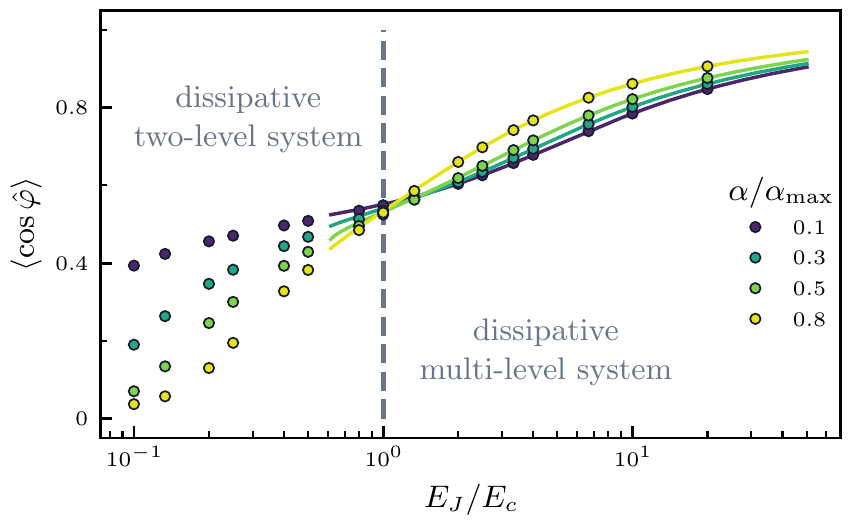}
    \vspace{-0.5cm}
    \caption{Josephson tunnelling $\big<\cos(\hat{\varphi})\big>$ in the ground state
of the charge-boson model, comparing full NRG simulation (dots) with the simple 
Ansatz~(\ref{eq:compactification}) (lines),
computed as a function of $E_J/E_c$ for offset charge $n_g=1/2$ and several values of
the normalized dissipation strength $\alpha/\alpha_\mathrm{max}$. Dissipation
tends to localize the phase for multilevel charge qubits ($E_J>E_c$), namely 
$\big<\cos(\hat{\varphi})\big>$ increases with $\alpha$, while conversely phase delocalizes 
for two-level qubits ($E_J<E_c$). Surprisingly, zero point fluctuations are nearly dissipation 
insensitive in the crossover regime $E_J\simeq E_c$.}
    \label{fig:NRGplot}
\end{figure}

In the absence of a quantum phase transition, one may be tempted to conclude that
the multilevel regime of dissipative qubits is trivial. On the contrary, it presents 
interesting many-body physics that we explore in the rest of this Letter.
We investigate zero point fluctuations of the superconducting phase
given by the average tunnelling $\big<\cos(\hat{\varphi})\big>$ in the
many-body ground state. For transmons ($E_J\gg E_c$), $\big<\cos(\hat{\varphi})\big>$
increases with dissipation $\alpha$, because the phase is damped by its environment
towards the minimum $\hat{\varphi}=0$ of the Josephson potential (this
behavior is not captured by a two-level approximation~\cite{SupInfo}).
In contrast, for a Cooper pair box ($E_c \gg E_J$) at the charge offset $n_g=1/2$, 
$\big<\cos(\hat{\varphi})\big>$
decreases with dissipation because charge fluctuations between $n=0$ and
$n=1$ tend to freeze, so that, due to the Heisenberg principle, the phase delocalizes.
This regime is strongly $n_g$ sensitive, since for $n_g=0$ the charge is already
frozen in absence of dissipation.
Both behaviors are clearly evidenced in Fig.~\ref{fig:NRGplot}, which shows
$\big<\cos(\hat{\varphi})\big>$ against $E_J/E_c$, for several values of the
normalized dissipation strength $\alpha/\alpha_\mathrm{max}$ at $n_g=0.5$
(dots are results from NRG simulations).
Remarkably, the crossover is characterized by quantum fluctuations
of the superconducting phase that are nearly dissipation-insensitive as seen by the
narrowing spread of the points at $E_J=E_c$. This striking behavior
is a manifestation of the frustrated nature of the qubit pointer states, that are neither 
purely phase-like nor purely charge-like in the crossover from multilevel to two-level qubits.


In order to capture physically these effects related to discrete charge, we finally develop 
a new description of dissipative multilevel qubits, since current polaronic theory~\cite{Silbey,bera_stabilizing_2014}
applies mainly to the two-level regime.
Obviously, dissipation tends to localize the phase in the multilevel regime, so that 
the wavefunction stays mosly trapped at the minima of the cosine Josephson potential.
For $\varphi^2\ll1$,
the self-consistent harmonic approximation (SCHA)~\cite{Zaikin,Giam,Leger:2019vv}
replaces the cosine potential in~(\ref{eq:fullH}) by an harmonic term:
\begin{equation}
\label{eq:linH}
\hat{H}_{\text{SCHA}} = \sum_k \omega_k \,\cre{k}\anh{k}
+ \hat{n}\sum_k ig_k(\cre{k}-\anh{k}) +
4E_c\hat{n}^2 + \frac{E_J^\star}{2}\hat{\varphi}^2,
\end{equation}
which is nothing but the Caldera-Leggett model of a damped harmonic oscillator
with renormalized Josephson energy $E_J^\star$.
However, charge discreteness has to be taken into account via the compactness of the phase 
$\varphi\in[0,2\pi]$,
which can be restored~\cite{mizel_right-sizing_2019,Bender}
by periodizing the vacuum of $\hat{H}_\text{SCHA}$ (denoted $\ket{0}_\text{SCHA}$):
\begin{equation}
    \ket{0_\circlearrowleft} =
    \sum_{w \in\Z} e^{i 2\pi w \hat{n}} e^{-in_g\hat{\varphi}} \ket{0}_\text{SCHA},
\label{eq:compactification}
\end{equation}
including a gate offset $n_g$ associated to Aharonov-Casher interference
\cite{bell_spectroscopic_2016}.
After diagonalizing the linear Hamiltonian~(\ref{eq:linH}) in eigenmodes $b^\dagger_\mu$,
$ \hat{H}_\text{SCHA} = \sum_\mu \Omega_\mu\big(\creb{\mu} \anhb{\mu} +
\nicefrac{1}{2} \big)$,
the qubit operators read $\hat{n}=i \sum_\mu v_\mu(b^\dagger_\mu-b_\mu)$ and
$\hat{\varphi}=\sum_\mu u_\mu(b^\dagger_\mu + b_\mu)$~\cite{SupInfo}.
Using $E_J^\star$ as variational parameter, we estimate the ground state energy
of Hamiltonian~(\ref{eq:fullH}), and obtain analytically the tunneling:
\begin{equation}
\label{cosvar}
\big<\!\cos(\hat{\varphi})\big>= \sum_{w \in \Z}\!
\left[\frac{(\pi w)^2}{2} \!+\! (-1)^w e^{-\frac{u^2}{2}} \right]\! e^{-2(\pi w)^2 v^2 - i2\pi w n_g},
\end{equation}
where $u^2 \equiv \sum_\mu u_\mu^2$, $v^2 \equiv \sum_\mu v_\mu^2$.
The lines in Fig.~\ref{fig:NRGplot} compare the full NRG simulations to the simple
formula~(\ref{cosvar}) in the range $E_J/E_c>1/2$, with excellent agreement. 
The crossover from multilevel to two-level regime results from a competition between 
two clear physical effects:
1) the Franck-Condon term $e^{-2 \pi^2 v^2 w^2}$ associated to the winding number $w$ 
dual to the qubit charge $n$, weighting the overlaps between wells;
2) the Aharonov-Casher phase $e^{- i2\pi w n_g}$ associated to the gate
charge $n_g$, driving interferences between wells.

In conclusion, we have demonstrated that realistic superconducting qubits do not show
the same dissipative properties predicted from models based on the two-level approximation. 
We also provided a new physical picture of the many-body wavefunction for dissipative multilevel
qubits, based on charge discreteness. Regarding experimental attempts
at simulating quantum spins with superconducting circuits, we found that reaching the spin-boson 
quantum phase transition requires very strong non-linearities, well beyond the transmon regime.
Similar considerations could apply to a wide class of model Hamiltonians that are touted
as candidates for quantum simulators ~\cite{FornDiaz:2016bo,Leger:2019vv,Joyez,Roy_SineGordonSimulation}.

\begin{acknowledgments}
KK and SB would like to thank IRCC, IITB for funding.
SB acknowledges support from SERB-DST, India, through Ramanujan Fellowship No.
SB/S2/RJN-128/2016,
Early Career Research Award No. ECR/2018/000876, Matrics No. MTR/2019/000566, and MPG
through the Max Planck Partner Group at IITB.
SF thanks hospitality of IITB and Wits University, and acknowledges
funding by PICS contract FermiCats.
\end{acknowledgments}

%

\vspace{3cm}
\onecolumngrid
\begin{center}
\textbf{\large Supplementary Materials for ``Spin-boson quantum phase transition in 
multilevel superconducting qubits''}
\end{center}
\setcounter{equation}{0}
\setcounter{figure}{0}
\setcounter{table}{0}
\setcounter{page}{1}
\makeatletter
\renewcommand{\theequation}{S\arabic{equation}}
\renewcommand{\thefigure}{S\arabic{figure}}
\renewcommand{\bibnumfmt}[1]{[S#1]}
\renewcommand{\citenumfont}[1]{S#1}

\section{Microscopic parameters for the charge-boson model}
In this section, we review one possible method for establishing the
microscopic Hamiltonian from the circuit model, Eq.~(4) of the main text.
The starting point is the Lagrangian
\begin{equation}
    \mathcal{L} = \frac{1}{2}\dot{\vec{\Phi}}^{\, \trsp}
\bm{C}\dot{\vec{\Phi}} - \frac{1}{2}\vec{\Phi}^{\, \trsp} \nicefrac{\bm
1}{\bm L}\,\vec{\Phi} + E_J \cos(\Phi_0) - n_g\dot{\Phi}_0,
\label{startL}
\end{equation}
written in units of $\hbar=2e=1$, and with the capacitance and inductance matrices :
\begin{eqnarray}
  \,\bm{C} = \begin{bmatrix}
      C_J+C_c & -C_c  &        &        & \\
      -C_c    & C+C_c+C_g & -C     &        & \\
              & -C        & 2C+C_g & -C     & \\
              &           & -C     & \ddots & \ddots \\
              &           &        & \ddots &
  \end{bmatrix},
  \quad {\nicefrac{\bm 1}{\bm L}} = \frac{1}{L} \begin{bmatrix}
      0 & 0  &    &       & \\
      0 & 1  & -1 &       & \\
        & -1 & 2  & -1    & \\
        &    & -1 & \ddots& \ddots \\
        &    &    & \ddots&
  \end{bmatrix}.
\end{eqnarray}
We denoted here $C_J=C_s+C_g$ the combination of the shunting capacitance $C_s$ and ground
capacitance $C_g$ of the qubit (see Fig.~1 of the main text).

\textbf{Diagonalisation.}
A half-infinite chain is here assumed. We want to find a basis where both capacitance and inductance
quadratic forms are diagonal. Since $\bm{C}$ is positive definite, this is equivalent to solving
$\nicefrac{\bm 1}{\bm L}\,\bm{P} = \bm{CP\omega}^2$ (a generalized eigenvalue problem). Under the
change of basis $\vec{\Phi} = \bm{P} \vec{\phi}$, the Lagrangian is :
\begin{equation}\label{eq:lagrangian2}
    \mathcal{L} = \frac{1}{2}\dot{\vec{\phi}}^{\, \trsp} \bm{P}^\trsp
\bm{CP}\dot{\vec{\phi}} - \frac{1}{2}\vec{\phi}^{\, \trsp}\bm{P}^\trsp
\bm{CP\omega}^2\vec{\phi}+ E_J \cos\left(\sum_l P_{0l}\phi_l\right) - n_g\sum_l P_{0l}\dot{\phi}_l.
\end{equation}
$\bm{P}^\trsp \bm{CP}$ is diagonal, because it commutes with $\bm{\omega}^2$. We
can then scale the eigenvectors to have $\bm{P}^\trsp \bm{CP}= \openone$, and
we reach Eq.~(2) of the main text. Written explicitly, the generalized eigenvalue problem gives :
\begin{align}
    \text{site }i=0: \quad & P_{1l} = \frac{C_J+C_c}{C_c}P_{0l}, \label{eqSupp:site0}\\
    \text{site }i=1: \quad & P_{2l} = P_{1l} \left(1 -
\frac{C_f}{C}\frac{\omega_{k_l}^2}{\omega_0^2-\omega_{k_l}^2} \right),
    \quad \text{with} \quad C_f = \frac{C_c C_J}{C_c+C_J} +C_g
    \quad \text{and} \quad \omega_0=  \frac{1}{\sqrt{LC}}, \label{eqSupp:site1}\\
    \text{site }i>1: \quad & P_{i+1\,l}+P_{i-1\,l} =
    P_{il}\left(2-\frac{C_g}{C}\frac{\omega_{k_l}^2}{\omega_0^2-\omega_{k_l}^2} \right), \label{eqSupp:sitei}
\end{align}
where we parametrize the wavevector $k_l=\pi l/N_\mathrm{modes}$ of mode
number $l$, with $N_\mathrm{modes}$ the number of modes (or sites in the chain).
The eigenproblem is almost invariant by translation, except for the boundary condition.
We assume that the solution obeys the form : $P_{jl} = N_l
\cos{\left((j-1)k_l + \theta_l\right)}$, with $N_l$ the normalisation factor, and $\theta_l$ a phase
shift due to the boundary. Note that $P_{0l}$ follows instead condition~\eqref{eqSupp:site0}, and
therefore is not part of the parametrization, hence the `$(j-1)$' labeling of the sites.
In the half-infinite chain limit $N_\mathrm{modes}\to\infty$, the wavenumbers continuously fill
the Brillouin zone : $k_l \in [0, \pi[$. The dispersion relation is obtained by injecting the
solution in the bulk equation~\eqref{eqSupp:sitei}, resulting in the standard
expression:
\begin{equation} \label{eq:dispersion}
  \omega_{k_l}^2 = \frac{4}{LC_g}
\frac{\sin^2(k_l/2)}{1+4\left(C/C_g\right)\sin^2(k_l/2)}.
\end{equation}
Besides the dispersion relation, we also need $P_{0l}$, which appears in the
coupling term between the charge qubit and the modes of the chain. The phase shift $\theta_l$ is
imposed by the boundary equation~\eqref{eqSupp:site1}, used together with
dispersion relation, and reads:
\begin{equation}
    \tan\theta_l = \tan \left(\frac{k_l}{2}\right) \left(2\frac{C_f}{C_g}-1 \right).
\end{equation}
Finally, we have to compute the normalization factors. First, one should note
that the  $k=0$ eigenfrequency is 0.
Equations~\eqref{eqSupp:site0},~\eqref{eqSupp:site1},~\eqref{eqSupp:sitei} do
not hold in this case, such that matrix elements $P_{i0}$ must be computed by
normalization of  $\bm{P}^\trsp \bm{CP}$, which leads to
$P_{00}=1/\sqrt{C_J+C_c}, \quad P_{i0}=0 \; \forall i>0$. This zero mode is then localized
on the zeroth site: we recognize the qubit degree of freedom, which will be singled
out as in the main text by a change of variable. The other
normalisation factors are computed from the following matrix elements :
\begin{align}
    \forall l, l' \neq 0, \quad \sum_{i,j=1}^\infty P_{il}C_{ij}P_{jl'}
    = \left(C_g \sum_{i=1}^\infty P_{il}P_{il'}
    + \frac{C_cC_J}{C_c+C_J}P_{1l}P_{1l'}\right)\left(1+4\frac{C}{C_g}\sin^2 (k_l/2)\right).
\end{align}
This matrix must be diagonal, so we collect only the terms proportional
to $\delta_{l,l'}$. Expanding the first part as
\begin{align}
    \sum_{i=1}^\infty P_{il}P_{il'}
    = \frac{1}{2}N_lN_{l'} \sum_{j=0}^{\infty}\quad &\cos(j(k_l+k_{l'}))\cos(\theta_l+ \theta_{l'})
    + \cos(j(k_l-k_{l'}))\cos(\theta_l- \theta_{l'}) \nonumber \\
    -&\sin(j(k_l+k_{l'}))\sin(\theta_l+ \theta_{l'})+ \sin(j(k_l-k_{l'}))\sin(\theta_l- \theta_{l'}).
\end{align}
Using the identity $\lim_{n\rightarrow \infty}\sum_{j=0}^n \cos(j k_l) =
\nicefrac{1}{2} + n\delta_{l,0}$
we can deduce the normalisation factor $N_l$.
With equation~\eqref{eqSupp:site0}, we get an analytic expression of the couplings for a
large but finite number of modes $N_\mathrm{modes}$:
\begin{equation}
    P_{00}=\frac{1}{\sqrt{C_J+C_c}}, \quad
P_{0l}=\frac{C_c}{C_J+C_c}\sqrt{\frac{2}{N_\mathrm{modes}}}
    \left( C_g+4C \sin^2(k_l/2) \right)^{-\frac{1}{2}}
    \left( 1+\left( 2\frac{C_f}{C_g}-1\right)^2\tan^2(k_l/2)\right)^{-\frac{1}{2}}.
\end{equation}
These expressions match the results from the numerical diagonalization of the Lagrangian~(\ref{startL}) performed with a finite number $N_\mathrm{modes}$ of sites, as soon as
$N_\mathrm{modes}\gtrsim 10$.

\textbf{Hamiltonian form.} Once in diagonal form, the Lagrangian \eqref{eq:lagrangian2} reads
\begin{equation}
    \mathcal{L} = \frac{1}{2}\sum_k\left(\dot{\phi}_k^2 - \omega_k^2\phi_k^2 \right)
    +E_{\rm J} \cos\left(\sum_k P_{0k}\phi_k \right) - n_g \sum_k P_{0k} \dot{\phi}_k.
\end{equation}
The Hamiltonian expression is obtained by Legendre transformation, using conjugate momenta
$N_k = \partial \mathcal{L}/\partial \dot{\phi_k}$:
\begin{equation}
    \mathcal{H} = \sum_k N_k \dot{\phi}_k - \mathcal{L}
    = \frac{1}{2}\sum_k\left((N_k+n_gP_{0k})^2 + \omega_k^2 \phi_k^2 \right)
    -E_{\rm J}\cos\left(P_{0k}\phi_k\right).
\end{equation}
Canonical quantization is now straightforward : all dynamical variables are promoted to
operators, obeying the commutation rule $[\phi_k, N_l] = i\delta_{kl}$. In this Hamiltonian
expression, the qubit degree of freedom does not appear explicitly. It can be reinstated
following a change of variables that conserves the commutation rules:
\begin{equation}
    \left\{ \begin{array}{l}
        \varphi = \sum_k P_{0k}\phi_k \\
        \varphi_m = \phi_m
    \end{array} \right.,
    \left\{ \begin{array}{l}
        n = N_0/P_{00} \\
        n_m = N_m- (P_{0m}/P_{00}) N_0
    \end{array} \right.,
\end{equation}
where $\vec{n}$ (resp. $\vec{N}$) is the vector of charges conjugate to
$\vec{\varphi}$ (resp. $\vec{\phi}$).
As a result, we obtain the Hamiltonian:
\begin{align}
    \hat{H} = (P_{00}^2+ \sum_m P_{0m}^2) (\hat{n}-n_g)^2 + 2 (\hat{n}-n_g) \sum_m P_{0m} \hat{n}_m
    + \frac{1}{2} \sum_{m}\left(\hat{n}_m^2 + \omega_m^2\hat{\varphi}_m^2 \right)
    - E_{\rm J} \cos(\hat{\varphi}).
\end{align}
It is noteworthy that the change of variables does not complicate this expression,
thanks to the fact that $\omega_0=0$ (this is a general feature of our model, because
the qubit mode does not participate in the inductance matrix ${\bf L}$). Otherwise, 
$\omega_0^2 \hat{\phi}_0^2$ 
would have transformed into:
\begin{equation}
    \omega_0^2 \hat{\phi}_0^2 = \left(\frac{\omega_0}{P_{00}}\right)^2\hat{\varphi}^2
    + \left(\frac{\omega_0}{P_{00}}\right)^2\left(\sum_m P_{0m}\hat{\varphi}_m\right)^2
    - 2\left(\frac{\omega_0}{P_{00}}\right)^2 \hat{\varphi} \sum_m{\hat{\varphi_m}},
\end{equation}
the right-hand side terms being respectively interpreted as qubit inductive energy, a diamagnetic 
`$A^2$' term, and a supplementary coupling between the qubit and the array. If present, these terms
would violate phase compactness. Finally, the array normal modes are expressed in terms of 
creation/annihilation operators, defined by:
\begin{equation}
            \hat{\varphi}_m = \frac{1}{\sqrt{2\omega_m}} (\cre{m} + \anh{m})
            , \quad \hat{n}_m = i \sqrt{\frac{\omega_m}{2}} (\cre{m} - \anh{m})
            \quad \text{and} \quad [\anh{n}, \cre{m}] = \delta_{nm}.
\end{equation}

\textbf{Spectral density.} The bath spectral density gives a more convenient tool to describe
this system with a small number of relevant parameters. It is defined as a continuous
function of frequency, $J(\omega) =  \pi \sum_l g_{k_l}^2 \delta(\omega - \omega_{k_l})$.
The $g_k$ couplings are defined in the main text as $g_{k_l} = \sqrt{\omega_{k_l}/2}P_{0l}$.
All the dependencies in the wave number $k_l$ can be expressed in terms of $\omega_k$ using
the dispersion relation \eqref{eq:dispersion}. We also change sums over modes to
integrals in the limit $N_\mathrm{modes}\to \infty$:
\begin{align}
& \frac{1}{N_\mathrm{modes}} \sum_l F[k_l] \to \frac{1}{\pi} \int_0^\pi \!\!\!
\mathrm{d}k \; F[k],\\
    &J(\omega) =  N_\mathrm{modes} \ddroit{k}{\omega} [g(\omega)]^2
    = \left(\frac{C_c}{C_c + C_J} \right)^2 \sqrt{\frac{L}{C_g}} \ \omega\
        \frac{\sqrt{1-\omega^2/\omega_{ P}^2}}
        {1+\omega^2\left(1/\omega_{Q}^2 - 1/\omega_{P}^2\right)}\, \theta(\omega_P - \omega),
\label{Seq:J}
\end{align}
with $\omega_{P} = \omega_0/\sqrt{1+4C/C_g}$ the plasma
frequency of the chain, and
$\omega_{Q} = \omega_{P}
\sqrt{1+\nicefrac{4C}{C_g}}/(\nicefrac{2C_f}{C_g}-1)$ a characteristic frequency
related to the qubit. From Eq.~(\ref{Seq:J}), it is clear that the denominator is
never vanishing within the band $\omega\in[0,\omega_P]$ (otherwise $J(\omega)$
would be singular). Using the expression for
$\omega_P$ and $\omega_Q$, and $1/\omega_J^2\equiv1/\omega_Q^2-1/\omega_P^2$, we
recover the low frequency cutoff of the qubit $\omega_J$ defined in the main
text.

For small frequencies, $J(\omega)$ obeys the so-called Ohmic behavior, $J(\omega) \simeq 2 \pi \alpha \omega$,
which defines the coupling strength $\alpha$. A higher frequencies, $J(\omega)$
quickly vanishes as $1/\omega^2$, provided $\omega_{J} \ll \omega_{P}$.
Most experimental devices verify this condition.
On the other hand, if $\omega_{J} \gg \omega_{P}$, $J(\omega)$ displays
a square-root hard cut-off at $\omega = \omega_{P}$.
In many cases, the exact form of the cut-off is not relevant, and we replace it by an exponential cut-off
at $\omega_{c} = \min \left\{\omega_{J}, \omega_{P} \right\}$,
$J(\omega) = 2 \pi \alpha \omega \exp(-\omega/\omega_{c})$, which is the spectral function used in the main
text to perform the numerical computations.

\textbf{Microscopic derivation of the electrostatic bound on dissipation.}
The charging energy of our microscopic circuit explicitly reads
\begin{equation}
E_c=\frac{1}{8}\left(C_J+C_c-\frac{C_c^2}{C_c+\frac{C_g}{2}+\sqrt{C_g\left(\frac{C_g}{4}+C\right)}}\right)^{-1}.
\end{equation}
This result can be obtained by noting that $E_c=\bm{C}^{-1}_{0 0}/8$ and analytically inverting the capacitance matrix. From equation~(5) in the main
text one can reach the same result, recast into
\begin{equation}
8E_c=\frac{1}{C_c+C_J}
    +2\pi\alpha\frac{\omega_J^2}{\omega_P}\left(\sqrt{1+\frac{\omega_P^2}{\omega_J^2}}-1\right)
\geqslant 2\pi\alpha\frac{\omega_J^2}{\omega_P}\left(\sqrt{1+\frac{\omega_P^2}{\omega_J^2}}-1\right),
\end{equation}
by integrating the spectral function \eqref{Seq:J} over $\omega$. Thus the dissipation strength $\alpha$ obeys the inequality
\begin{equation}
\alpha \leqslant \frac{4E_c}{\pi\omega_P}\left(\sqrt{1+\frac{\omega_P^2}{\omega_J^2}}+1\right).
\end{equation}
When $\omega_J \ll \omega_P$ (which is the typical situation for realistic
devices), the electrostatic bound $\alpha \leqslant 4 E_c/(\pi \omega_J)$
assumes the same form as in the main text (albeit with a different numerical
prefactor), while for $\omega_J \gg \omega_P$ the bound reads
$\alpha \leqslant 8 E_c/(\pi \omega_P)$.

\section{Compact ansatz wavefunction for charge sensitive circuits}
We build in this section the compact ansatz step by step, following the outline
of the main text. While our approach is completely generic to charge
sensitive superconducting circuits, we focus here on the charge-boson
Hamiltonian:
\begin{equation}
    \hat{H}
    = 4E_c(\hat{n}-n_g)^2 - E_J\cos\hat{\varphi} + (\hat{n}-n_g) \sum_{k}ig_k (\hat{a}^\dag_k - \hat{a}_k)
+ \sum_k \omega_k \hat{a}^\dag_k \hat{a}_k,
\label{eqSupp:Hchargeboson}
\end{equation}
where the discrete sums over the wave vector $k$ run on the Brillouin zone
$[0,\pi]$.
The renormalized linear approximation, that holds only in the deep transmon regime where the phase
fluctuations are much smaller that $2\pi$, is used as a linearized parent Hamiltonian for our
variational trial state. It reads:
\begin{equation}
    \hat{H}_\mathrm{SCHA} = 4E_c\hat{n}^2 + \frac{E_J^\star}{2} \hat{\varphi}^2 + i\hat{n} \sum_{k}g_k (\hat{a}^\dag_k - \hat{a}_k) + \sum_k \omega_k \left(\hat{a}^\dag_k \hat{a}_k+\nicefrac{1}{2}\right).
\label{eqSupp:Hlinear}
\end{equation}
Here the charge offset $n_g$ was gauged out since the phase is uncompact in
the linear approximation, and $E_J^\star$ is a free parameter used
in the variational method, after compactification is applied.
This Hamiltonian can be brought to diagonal form by a
Bogoliubov rotation mixing the qubit degree of freedom and bosons from the
environment. For this purpose, we introduce rescaled charge $\hat{n}_k$ and
phase $\hat{\phi}_k$ normal modes, so that $\hat{a}_k = (i\omega_k \hat{n}_k 
+ \hat{\phi}_k)/\sqrt{2\omega_k}$. We then lump these normal modes together with the 
qubit degree of freedom in the set charges 
$\hat{n}_\mu = (\hat{n}, \hat{n}_1, \hat{n}_2, \hdots)$ and phases
$\hat{\phi}_\mu = (\hat{\varphi}, \hat{\phi}_1, \hat{\phi}_2, \hdots)$
(with greek indices), so
that Hamiltonian~(\ref{eqSupp:Hlinear}) reads:
\begin{equation}
\hat{H}_\mathrm{SCHA} = \frac{1}{2} \sum_\mu \hat{\phi}_\mu^2
+ \frac{1}{2} \sum_{\sigma\rho}\hat{n}_\sigma M_{\sigma \rho} \hat{n}_\rho \\
\qquad \text{where} \;\bm{M}=
    \begin{bmatrix}
    8E_J E_c & g_1\sqrt{2 E_J \omega_1}&g_2\sqrt{2 E_J \omega_2}& \hdots \\
    g_1\sqrt{2 E_J \omega_1} & \omega_1^2 &            &  \\
    g_2\sqrt{2 E_J \omega_2} &            & \omega_2^2 &  \\
     \vdots                  &            &            & \ddots
    \end{bmatrix} . 
\label{eqSupp:Mmatrix}
\end{equation}
Diagonalization of the matrix $\bm{M}$,
which assumes an arrowhead form, can be done efficiently from dedicated algorithms, 
as well as perturbative series expansion. We obtain $\bm{M}=\bm{O} \bm{D} \bm{O}^{\, \trsp}$,
with an orthonormal $\bm{O}$ matrix 
and a diagonal matrix $D_{\mu\nu} = \Omega_{\mu}^2 \delta_{\mu\nu}$, introducing $\Omega_\mu$ 
the eigenfrequencies of the linear system.
Defining new conjugate variables $\tilde{n}_\mu = \sum_\nu O_{\mu\nu} \hat{n}_\nu$ and 
$\tilde{\phi}_\mu = \sum_\nu O_{\mu\nu} \hat{\phi}_\nu$, we get:
\begin{equation}
\hat{H}_\mathrm{SCHA} = \frac{1}{2} \sum_\mu \left[ \tilde{\phi}_{\mu}^2 
+ \Omega_{\mu}^2 \tilde{n}_{\mu}^2\right] = \sum_\mu \Omega_\mu 
\left[\creb{\mu}\anhb{\mu}+ \frac{1}{2}\right]
\end{equation}
where we introduced the destruction operators $\tilde{b}_\mu = (i \Omega_\mu \tilde{n}_\mu 
+ \tilde{\phi}_\mu)/\sqrt{2\Omega_\mu}$.

In order to perform the compactification of the qubit phase $\hat\varphi$, we
express the qubit charge in terms of the final eigenmodes of $H_\mathrm{SCHA}$:
\begin{eqnarray}
\hat{n} &=& \hat{n}_0 = \sum_\mu O_{\mu0}\tilde{n}_\mu 
= \sum_\mu O_{\mu0}\sqrt{\frac{1}{2\Omega_\mu}} (\creb{\mu} - \anhb{\mu})
\equiv i \sum_\mu v_\mu(\creb{\mu} - \anhb{\mu}) \\
\hat{\varphi} &=& \hat{\varphi}_0 = \sum_\mu O_{\mu0}\tilde{\varphi}_\mu 
= \sum_\mu O_{\mu0}\sqrt{\frac{\Omega_\mu}{2}} (\creb{\mu}+\anhb{\mu})
\equiv \sum_\mu u_\mu (\creb{\mu}+\anhb{\mu}). \label{eqSupp:decompo}
\end{eqnarray}
We then enforce the periodic boundary conditions (compactification) by
repeatedly displacing the ground state of $H_{\rm SCHA}$, noted $\ket{0}$, by an
integer times $2\pi$:
\begin{equation}
    \ket{0_\circlearrowleft}=\sum_{w \in \Z} e^{i2\pi w \hat{n}}\ket{0}
= \sum_{w \in \Z} e^{-2\pi w \sum_\mu v_\mu (\creb{\mu} - \anhb{\mu})}\ket{0} , 
\quad \text{where } \anhb{\mu}\ket{0} 
= 0 \quad \forall \mu.
\label{eqSupp:compactWF}
\end{equation}
Standard coherent state algebra in terms of the normal modes $\anhb{\mu}$ allows to readily
compute expectations values from the compactified state~(\ref{eqSupp:compactWF}).

\textbf{Regularisation.} With such a definition, $\ket{0_\circlearrowleft}$ has
infinite norm, because the associated wave function is both periodic and defined
over $\R$. Indeed, by re-indexing sums over winding numbers,
\begin{equation}
    \bracket{0_\circlearrowleft}{0_\circlearrowleft}
    = \sum_{v, w \in \Z} \int_\R \di{\varphi} \bracket{0}{\varphi - 2\pi v} \bracket{\varphi + 2\pi w}{0}
    = \bracket{0}{0_\circlearrowleft}\left( \sum_{v \in \Z} 1 \right),
\end{equation}
which is clearly infinite. A way out is to restrict the wave-function over the
interval $[0, 2\pi [$, which is in fact equivalent to simply drop the infinite
factor in the last expression:
\begin{align}
    \bracket{0}{0_\circlearrowleft}\left( \sum_{v \in \Z} 1 \right) \xrightarrow[\text{to }[0, 2 \pi[ ]{\text{restricted}}
    & \sum_{v, w \in \Z} \int_0^{2\pi} \di{\varphi} \bracket{0}{\varphi-2 \pi v }\bracket{\varphi + 2\pi w}{0} \nonumber \\
    &= \sum_{v, w \in \Z} \int_{2\pi v}^{2\pi(v+1)} \di{\varphi} \bracket{0}{\varphi}\bracket{\varphi+2\pi(v+w)}{0}
    = \bracket{0}{0_\circlearrowleft}.
\end{align}
The last line is obtained with a relabeling of the sums $w'=w+v$, and patching
the integrals together to get back an integral on $\R$. The same trick can be
used for expectation values of any operator $\hat{\mathcal{O}}$, provided that
it is itself $2\pi$-periodic, which means that $[\hat{\mathcal{O}}, \sum_w
\exp(i 2 \pi w \hat{n})] =0$. \\

\textbf{Aharonov-Casher phases.} We already mentioned that $n_g$ acts as a gauge
potential on the system. It can usually be removed from the Hamiltonian
by a gauge transformation
$\hat{U} = \exp{(i n_g \hat{\varphi})}$, which however affects the boundary
condition on the phase (unless the model is not compact).
Under such a transformation,
$\hat{H}_\mathrm{SCHA}(n_g) = \hat{U}^\dag\hat{H}_\mathrm{SCHA}(0)\hat{U}$. It can
be checked that $\hat{U}^\dag \ket{0}$ is an eigenstate of
$\hat{H}_\mathrm{SCHA}(n_g)$. For a non-compact model, the gauge has no observable
effect, but the compactification process will change this state of affairs, since
\begin{equation}
    \sum_{w \in \Z}e^{i 2 \pi w \hat{n}} \hat{U}^\dag \ket{0}
    = \hat{U}^\dag \sum_{w  \in \Z} e^{i 2 \pi w(\hat{n}-n_g)}\ket{0}.
\end{equation}
As an example, the effect on the ansatz norm is :
\begin{equation}
    \bracket{0_\circlearrowleft}{0_\circlearrowleft}
    = \sum_{v, w \in \Z} \bra{0} e^{-i2\pi v (\hat{n}-n_g)}\hat{U}\hat{U}^\dag e^{i2\pi w(\hat{n}-n_g)}\ket{0}
    = \sum_{w \in \Z}  e^{-i2\pi w n_g}\bra{0}e^{i2\pi w\hat{n}}\ket{0},
\end{equation}
using $\hat{U}\hat{U}^\dag = \openone$ and the same infinite factor canceling
argument as before. The offset charge effect is seen as an Aharonov-Casher phase
that depends on the winding number. The interference between different winding
numbers will create an observable effect due to the gauge. The same argument can
be used for the expectation value of any gauge invariant operator
$\hat{\mathcal{O}}$, \textit{i.e.} $[\hat{\mathcal{O}}, \hat{U}]=0$. \\

\begin{figure}[htb]
    \centering
    \begin{minipage}{0.7 \textwidth}
    \includegraphics{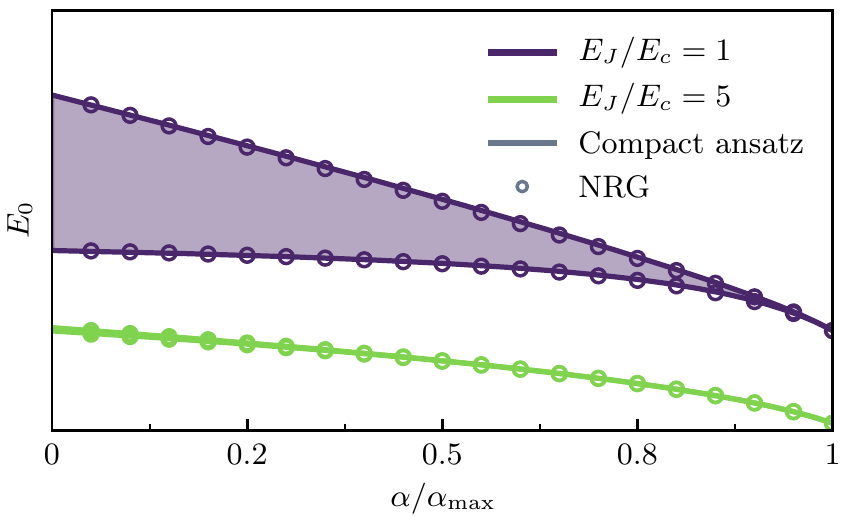}
    \end{minipage}
    \begin{minipage}{0.29 \textwidth}
    \caption{Ground state energy bands associated to the offset charge $n_g$,
both for a multi-level charge qubit (narrower band at the bottom, at $E_J/E_c=5$) and the crossover
regime (broaded band at the top, at $E_J/E_c=1$). The analytical expression~(\ref{eqSupp:energyExp})
from the compact Ansatz (lines) compares quantitatively to the full NRG simulation (dots).}
    \vfill
    \end{minipage}
    \label{figSupp:Energy}
\end{figure}

$\mathbf{E_{\it J}^\star}$ \textbf{optimisation.} The anharmonicity of the
$\cos$-shaped potential tends to soften the phase confinement compared to
quadratic potential, thus enhancing the zero point phase fluctuations. Having
built an ansatz adapted to the specifics of the problem, we can use it as
starting point for a variational method. The free parameter is the effective
stiffness of the potential $E_J^\star$ in the linearized
Hamiltonian~(\ref{eqSupp:Hlinear}).
We need to compute the energy expectation value of the full Hamiltonian~(\ref{eqSupp:Hchargeboson})
within the compactified ground state of the linearized Hamiltonian~(\ref{eqSupp:Hlinear}).
It is noteworthy that $\hat{H}$ is $2 \pi$-periodic, but not
gauge invariant. Instead, we make use of $\hat{U}\hat{H}(n_g) \hat{U}^\dag
=\hat{H}(0)$. Using once again the decomposition~\eqref{eqSupp:decompo},
\begin{align}
    \frac{\bra{0_\circlearrowleft}\hat{H}\ket{0_\circlearrowleft}}{\bracket{0_\circlearrowleft}{0_\circlearrowleft}}
    =\sum_\mu \frac{\Omega_\mu }{2}-E_J\frac{u^2}{2}
    - E_J \left (\sum_{w \in \Z} e^{-2(\pi w)^2v^2-i2\pi wn_g}\right)^{-1}\sum_{w \in \Z}
\left(\frac{(\pi w)^2}{2}+(-1)^w e^{-\frac{u^2}{2}} \right) e^{-2(\pi w)^2v^2-i2\pi wn_g}. \label{eqSupp:energyExp}
\end{align}
The $E_J^\star$ dependence is contained in $u^2$ and $v^2$. The full numerical
procedure consists, at every step, in a minimization of the
expression~(\ref{eqSupp:energyExp}) over $E_J^\star$,
using a fast diagonalization the arrowhead $\bm{M}$ matrix~\eqref{eqSupp:Mmatrix} for the new
value of $E_J^\star$, and the computation
of the two scalars $u^2$ and $v^2$. We then compute the energy expectation
value~\eqref{eqSupp:energyExp} with a number of terms in the sums over $w$
controlled by $v^2$. Since the $w\neq 0$ terms are exponentially suppressed, the
sums are rapidly convergent. In practice, we need at most $w \sim 10$ terms when
the Josephson energy is close to the breaking point of the method, $E_J/E_c \sim 1$.
Crucially, $u^2$ is independent of the number of modes. Overall, the complexity
of the whole procedure is $\mathcal{O}(N_{\text{modes}}^2)$, with
$N_{\text{modes}}$ the total number of modes in the chain. The quality of the Ansatz
is found to be excellent, see Fig.~\ref{figSupp:Energy} for a comparison of the
ground state energy obtained in the full NRG simulation.

\textbf{Zero point phase fluctuations.} Since $\cos \hat{\varphi}$ is both $2\pi$-periodic and gauge invariant, its expectation value is evaluated using the two previous tricks. Then, expressing every operator in terms of $\creb{\mu}, \anhb{\mu}$ with~\eqref{eqSupp:decompo} :
\begin{align}
    \bra{0_\circlearrowleft}\cos\hat{\varphi}\ket{0_\circlearrowleft}
    &= \frac{1}{2}\sum_{\pm, w\in\Z}e^{-i2\pi wn_g}
\bra{0}\exp\Big(\pm i \sum_\sigma u_\sigma(\creb{\sigma}+\anhb{\sigma})\Big)
\exp\Big(-2\pi w \sum_\rho v_\rho(\creb{\rho} - \anhb{\rho})\Big)\ket{0} \nonumber \\
    &= e^{-\frac{1}{2} \sum_\nu (u_\nu)^2}\sum_w (-1)^w
	e^{-2(\pi w)^2\sum_\mu (v_\mu)^2 - i 2\pi w n_g}.
\end{align}
Note that since the state norm isn't unity, normalization is necessary, by a
factor $\bracket{0_\circlearrowleft}{0_\circlearrowleft}=\sum_w
\exp(-2(\pi w)^2v^\mu v_\mu)$. Clearly, $v^2 \equiv \sum_\mu (v^\mu)^2$ weights the
corrections from non-zero winding numbers. It vanishes when $E_J \rightarrow
\infty$, providing a pure harmonic oscillator behavior in this limit. At finite
$E_J$, it sets the number of windings taken into account to reach required
numerical accuracy. In the same fashion, $u^2\equiv \sum_\mu (u^\nu)^2$ renormalizes
the bare Josephson energy, $E_J' = E_J \exp(-u^2/2)$ (note that the
previously defined term $E_J^\star$ appears only in the linearized Hamiltonian
used to derive the ansatz, and acts only as a variational parameter). \\

\section{Perturbation theory at small coupling strength.}
One of the charge-boson model's striking features is the different responses of the junction's phase
fluctuations $\langle\cos \hat{\varphi} \rangle$ to coupling strength $\alpha$, depending of the
$E_J/E_c$ regime, as shown by Fig. 3 of the main text. Broadly speaking, the environment damps the
phase fluctuations of the dissipative multi-level charge qubits, but enhances those of the dissipative 
two-level system. As already emphasized, this behavior cast doubt on the two-level description of 
dissipative multi-level qubits.
Arguably, while this feature is correctly described by both NRG and our compact
ansatz, a simple perturbative analysis in $\alpha$ should already be able to discriminate between
these two regimes, and pin-point the break down of the two-level approximation.

\begin{figure}[htb]
    \centering
    \begin{minipage}{0.7 \textwidth}
    \includegraphics{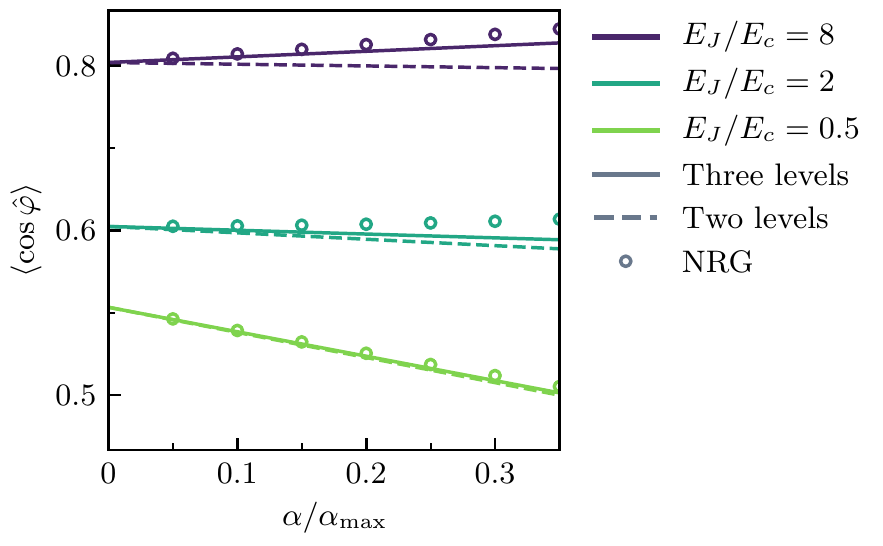}
    \end{minipage}
    \begin{minipage}{0.29 \textwidth}
    \caption{NRG estimates for phase fluctuations as a function of $\alpha$
    (circles), compared to the two and three levels approximation within first order
perturbation theory (dashed and solid lines respectively).
The two levels approximation (dashed lines) clearly fails for tmulti-level qubits,
where the slope changes of sign, while bringing the third level improves the
agreement to NRG. }
    \vfill
    \end{minipage}
    \label{figSupp:vsNRG}
\end{figure}

We employ time-independent perturbation theory at second order, with $(\hat{n} - \nicefrac{1}{2}) \sum_{k}ig_k (\hat{a}^\dag_k - \hat{a}^{\phantom{\dag}}_k)$ as the perturbation. We denote \ket{\psi_n} the eigenstates of the bare qubit, $E_n$ their energies. Then,
\begin{align}
    \avrg{\cos \hat{\varphi}} \simeq \bra{0}\cos \hat{\varphi} \ket{0}
    &+ \sum_{\substack{k\\ n \neq 0 \\ m \neq 0}}g_k^2 \frac{\bra{\psi_n}\hat{n}\ket{\psi_0}\bra{\psi_0}\hat{n}\ket{\psi_m}}
    {(E_0-E_n-\omega_k)(E_0-E_m-\omega_k)}\bra{\psi_n}\cos \hat{\varphi} \ket{\psi_m}
    - \sum_{\substack{k\\ n \neq 0}} g_k^2 \frac{|\bra{\psi_n}\hat{n}\ket{\psi_0}|^2}
    {(E_0-E_n-\omega_k)^2} \bra{\psi_0}\cos \hat{\varphi} \ket{\psi_0} \nonumber \\
    &+ 2\sum_{\substack{k, j, \\ i\neq 0}}g_k^2 \frac{\bra{\psi_0}\cos \hat{\varphi} \ket{\psi_i}\bra{\psi_j}\hat{n}\ket{\psi_0}\bra{\psi_j}\hat{n}\ket{\psi_i}}{(E_0-E_i)(E_0-E_J-\omega_k)}.
\end{align}
This expression takes into account the multi-level nature of the qubit. It can be reduced by restricting the sum on bare qubit levels to the most significant element:
\begin{align}
    \avrg{\cos \hat{\varphi}} \simeq \bra{0}\cos \hat{\varphi} \ket{0}
    &+ \sum_k g_k^2\frac{|\bra{\psi_1}\hat{n}\ket{\psi_0}|^2}
    {(E_0-E_1-\omega_k)^2}\bra{\psi_1}\cos \hat{\varphi}\ket{\psi_1}
    - \sum_k g_k^2\frac{|\bra{\psi_1}\hat{n}\ket{\psi_0}|^2}{(E_0-E_1-\omega_k)^2}
    \bra{\psi_0}\cos \hat{\varphi}\ket{\psi_0} \nonumber \\
    &+ 2\sum_k g_k^2 \frac{\bra{\psi_1}\hat{n}\ket{\psi_0}\bra{\psi_1}\hat{n}\ket{\psi_2}}
    {(E_0-E_2)(E_0-E_1-\omega_k)}\bra{\psi_2}\cos \hat{\varphi} \ket{\psi_0}.
\label{eqSupp:finalZPF}
\end{align}
The first and second term correspond to the two-level approximation (in the
limit $\alpha\to0$). However, the third term adds the contribution from the
third qubit level into the mix. Indeed, this term is mostly responsible for
the qualitative change between dissipative two- and multi-level qubits
when $E_J/E_c$ is increased, as shown by the Fig. \ref{figSupp:vsNRG}.

\begin{thebibliography}{65}%
\makeatletter
\providecommand \@ifxundefined [1]{%
 \@ifx{#1\undefined}
}%
\providecommand \@ifnum [1]{%
 \ifnum #1\expandafter \@firstoftwo
 \else \expandafter \@secondoftwo
 \fi
}%
\providecommand \@ifx [1]{%
 \ifx #1\expandafter \@firstoftwo
 \else \expandafter \@secondoftwo
 \fi
}%
\providecommand \natexlab [1]{#1}%
\providecommand \enquote  [1]{``#1''}%
\providecommand \bibnamefont  [1]{#1}%
\providecommand \bibfnamefont [1]{#1}%
\providecommand \citenamefont [1]{#1}%
\providecommand \href@noop [0]{\@secondoftwo}%
\providecommand \href [0]{\begingroup \@sanitize@url \@href}%
\providecommand \@href[1]{\@@startlink{#1}\@@href}%
\providecommand \@@href[1]{\endgroup#1\@@endlink}%
\providecommand \@sanitize@url [0]{\catcode `\\12\catcode `\$12\catcode
  `\&12\catcode `\#12\catcode `\^12\catcode `\_12\catcode `\%12\relax}%
\providecommand \@@startlink[1]{}%
\providecommand \@@endlink[0]{}%
\providecommand \url  [0]{\begingroup\@sanitize@url \@url }%
\providecommand \@url [1]{\endgroup\@href {#1}{\urlprefix }}%
\providecommand \urlprefix  [0]{URL }%
\providecommand \Eprint [0]{\href }%
\providecommand \doibase [0]{https://doi.org/}%
\providecommand \selectlanguage [0]{\@gobble}%
\providecommand \bibinfo  [0]{\@secondoftwo}%
\providecommand \bibfield  [0]{\@secondoftwo}%
\providecommand \translation [1]{[#1]}%
\providecommand \BibitemOpen [0]{}%
\providecommand \bibitemStop [0]{}%
\providecommand \bibitemNoStop [0]{.\EOS\space}%
\providecommand \EOS [0]{\spacefactor3000\relax}%
\providecommand \BibitemShut  [1]{\csname bibitem#1\endcsname}%
\let\auto@bib@innerbib\@empty
\bibitem [{\citenamefont {McArdle}\ \emph {et~al.}(2020)\citenamefont
  {McArdle}, \citenamefont {Endo}, \citenamefont {Aspuru-Guzik}, \citenamefont
  {Benjamin},\ and\ \citenamefont {Yuan}}]{QuantumComputationalChemistry}%
  \BibitemOpen
  \bibfield  {author} {\bibinfo {author} {\bibfnamefont {S.}~\bibnamefont
  {McArdle}}, \bibinfo {author} {\bibfnamefont {S.}~\bibnamefont {Endo}},
  \bibinfo {author} {\bibfnamefont {A.}~\bibnamefont {Aspuru-Guzik}}, \bibinfo
  {author} {\bibfnamefont {S.~C.}\ \bibnamefont {Benjamin}},\ and\ \bibinfo
  {author} {\bibfnamefont {X.}~\bibnamefont {Yuan}},\ }\href
  {https://doi.org/10.1103/RevModPhys.92.015003} {\bibfield  {journal}
  {\bibinfo  {journal} {Rev. Mod. Phys.}\ }\textbf {\bibinfo {volume} {92}},\
  \bibinfo {pages} {015003} (\bibinfo {year} {2020})}\BibitemShut {NoStop}%
\bibitem [{\citenamefont {Montanaro}(2016)}]{QuantumAlgorithms}%
  \BibitemOpen
  \bibfield  {author} {\bibinfo {author} {\bibfnamefont {A.}~\bibnamefont
  {Montanaro}},\ }\href@noop {} {\bibfield  {journal} {\bibinfo  {journal} {npj
  Quantum Information}\ }\textbf {\bibinfo {volume} {2}},\ \bibinfo {pages}
  {15023} (\bibinfo {year} {2016})}\BibitemShut {NoStop}%
\bibitem [{\citenamefont {Feynman}(1982)}]{FeynmanQuantum}%
  \BibitemOpen
  \bibfield  {author} {\bibinfo {author} {\bibfnamefont {R.}~\bibnamefont
  {Feynman}},\ }\href@noop {} {\bibfield  {journal} {\bibinfo  {journal} {Int.
  J. Theor. Phys.}\ }\textbf {\bibinfo {volume} {21}},\ \bibinfo {pages} {467}
  (\bibinfo {year} {1982})}\BibitemShut {NoStop}%
\bibitem [{\citenamefont {Ladd}\ \emph {et~al.}(2010)\citenamefont {Ladd},
  \citenamefont {Jelezko}, \citenamefont {Laflamme}, \citenamefont {Nakamura},
  \citenamefont {Monroe},\ and\ \citenamefont
  {O'Brien}}]{ReviewQuantumComputing}%
  \BibitemOpen
  \bibfield  {author} {\bibinfo {author} {\bibfnamefont {T.~D.}\ \bibnamefont
  {Ladd}}, \bibinfo {author} {\bibfnamefont {F.}~\bibnamefont {Jelezko}},
  \bibinfo {author} {\bibfnamefont {R.}~\bibnamefont {Laflamme}}, \bibinfo
  {author} {\bibfnamefont {Y.}~\bibnamefont {Nakamura}}, \bibinfo {author}
  {\bibfnamefont {C.}~\bibnamefont {Monroe}},\ and\ \bibinfo {author}
  {\bibfnamefont {J.~L.}\ \bibnamefont {O'Brien}},\ }\href
  {https://doi.org/10.1038/nature08812} {\bibfield  {journal} {\bibinfo
  {journal} {Nature}\ }\textbf {\bibinfo {volume} {464}},\ \bibinfo {pages}
  {45} (\bibinfo {year} {2010})}\BibitemShut {NoStop}%
\bibitem [{\citenamefont {Georgescu}\ \emph {et~al.}(2014)\citenamefont
  {Georgescu}, \citenamefont {Ashhab},\ and\ \citenamefont {Nori}}]{Nori_RMP}%
  \BibitemOpen
  \bibfield  {author} {\bibinfo {author} {\bibfnamefont {I.~M.}\ \bibnamefont
  {Georgescu}}, \bibinfo {author} {\bibfnamefont {S.}~\bibnamefont {Ashhab}},\
  and\ \bibinfo {author} {\bibfnamefont {F.}~\bibnamefont {Nori}},\ }\href
  {https://doi.org/10.1103/RevModPhys.86.153} {\bibfield  {journal} {\bibinfo
  {journal} {Rev. Mod. Phys.}\ }\textbf {\bibinfo {volume} {86}},\ \bibinfo
  {pages} {153} (\bibinfo {year} {2014})}\BibitemShut {NoStop}%
\bibitem [{\citenamefont {Kouwenhoven}\ and\ \citenamefont
  {Glazman}(2001)}]{Kouwenhoven:2001wd}%
  \BibitemOpen
  \bibfield  {author} {\bibinfo {author} {\bibfnamefont {L.}~\bibnamefont
  {Kouwenhoven}}\ and\ \bibinfo {author} {\bibfnamefont {L.}~\bibnamefont
  {Glazman}},\ }\href@noop {} {\bibfield  {journal} {\bibinfo  {journal}
  {Physics World}\ }\textbf {\bibinfo {volume} {14}},\ \bibinfo {pages} {33}
  (\bibinfo {year} {2001})}\BibitemShut {NoStop}%
\bibitem [{\citenamefont {Bloch}\ \emph {et~al.}(2008)\citenamefont {Bloch},
  \citenamefont {Dalibard},\ and\ \citenamefont {Zwerger}}]{Bloch:2008gl}%
  \BibitemOpen
  \bibfield  {author} {\bibinfo {author} {\bibfnamefont {I.}~\bibnamefont
  {Bloch}}, \bibinfo {author} {\bibfnamefont {J.}~\bibnamefont {Dalibard}},\
  and\ \bibinfo {author} {\bibfnamefont {W.}~\bibnamefont {Zwerger}},\ }\href
  {https://doi.org/10.1103/RevModPhys.80.885} {\bibfield  {journal} {\bibinfo
  {journal} {Reviews of Modern Physics}\ }\textbf {\bibinfo {volume} {80}},\
  \bibinfo {pages} {885} (\bibinfo {year} {2008})}\BibitemShut {NoStop}%
\bibitem [{\citenamefont {Houck}\ \emph {et~al.}(2012)\citenamefont {Houck},
  \citenamefont {T{\"u}reci},\ and\ \citenamefont {Koch}}]{Houck:2012iq}%
  \BibitemOpen
  \bibfield  {author} {\bibinfo {author} {\bibfnamefont {A.~A.}\ \bibnamefont
  {Houck}}, \bibinfo {author} {\bibfnamefont {H.~E.}\ \bibnamefont
  {T{\"u}reci}},\ and\ \bibinfo {author} {\bibfnamefont {J.}~\bibnamefont
  {Koch}},\ }\href {https://doi.org/10.1038/nphys2251} {\bibfield  {journal}
  {\bibinfo  {journal} {Nature Physics}\ }\textbf {\bibinfo {volume} {8}},\
  \bibinfo {pages} {292} (\bibinfo {year} {2012})}\BibitemShut {NoStop}%
\bibitem [{\citenamefont {Raftery}\ \emph {et~al.}(2014)\citenamefont
  {Raftery}, \citenamefont {Sadri}, \citenamefont {Schmidt}, \citenamefont
  {Tureci},\ and\ \citenamefont {Houck}}]{Raftery:2014jk}%
  \BibitemOpen
  \bibfield  {author} {\bibinfo {author} {\bibfnamefont {J.}~\bibnamefont
  {Raftery}}, \bibinfo {author} {\bibfnamefont {D.}~\bibnamefont {Sadri}},
  \bibinfo {author} {\bibfnamefont {S.}~\bibnamefont {Schmidt}}, \bibinfo
  {author} {\bibfnamefont {H.~E.}\ \bibnamefont {Tureci}},\ and\ \bibinfo
  {author} {\bibfnamefont {A.~A.}\ \bibnamefont {Houck}},\ }\href
  {https://doi.org/10.1103/PhysRevX.4.031043} {\bibfield  {journal} {\bibinfo
  {journal} {Physical Review X}\ }\textbf {\bibinfo {volume} {4}},\ \bibinfo
  {pages} {031043} (\bibinfo {year} {2014})}\BibitemShut {NoStop}%
\bibitem [{\citenamefont {Forn-D{\'\i}az}\ \emph {et~al.}(2017)\citenamefont
  {Forn-D{\'\i}az}, \citenamefont {Garcia-Ripoll}, \citenamefont {Peropadre},
  \citenamefont {Orgiazzi}, \citenamefont {Yurtalan}, \citenamefont
  {Belyansky}, \citenamefont {Wilson},\ and\ \citenamefont {Lupa{\c
  s}cu}}]{FornDiaz:2016bo}%
  \BibitemOpen
  \bibfield  {author} {\bibinfo {author} {\bibfnamefont {P.}~\bibnamefont
  {Forn-D{\'\i}az}}, \bibinfo {author} {\bibfnamefont {J.~J.}\ \bibnamefont
  {Garcia-Ripoll}}, \bibinfo {author} {\bibfnamefont {B.}~\bibnamefont
  {Peropadre}}, \bibinfo {author} {\bibfnamefont {J.~L.}\ \bibnamefont
  {Orgiazzi}}, \bibinfo {author} {\bibfnamefont {M.~A.}\ \bibnamefont
  {Yurtalan}}, \bibinfo {author} {\bibfnamefont {R.}~\bibnamefont {Belyansky}},
  \bibinfo {author} {\bibfnamefont {C.~M.}\ \bibnamefont {Wilson}},\ and\
  \bibinfo {author} {\bibfnamefont {A.}~\bibnamefont {Lupa{\c s}cu}},\ }\href
  {https://doi.org/10.1038/nphys3905} {\bibfield  {journal} {\bibinfo
  {journal} {Nature Physics}\ }\textbf {\bibinfo {volume} {13}},\ \bibinfo
  {pages} {39} (\bibinfo {year} {2017})}\BibitemShut {NoStop}%
\bibitem [{\citenamefont {Roushan}\ \emph {et~al.}(2017)\citenamefont {Roushan}
  \emph {et~al.}}]{Roushan:2017ij}%
  \BibitemOpen
  \bibfield  {author} {\bibinfo {author} {\bibfnamefont {P.}~\bibnamefont
  {Roushan}} \emph {et~al.},\ }\href {https://doi.org/10.1126/science.aao1401}
  {\bibfield  {journal} {\bibinfo  {journal} {Science}\ }\textbf {\bibinfo
  {volume} {358}},\ \bibinfo {pages} {1175} (\bibinfo {year}
  {2017})}\BibitemShut {NoStop}%
\bibitem [{\citenamefont {Ma}\ \emph {et~al.}(2019)\citenamefont {Ma},
  \citenamefont {Saxberg}, \citenamefont {Owens}, \citenamefont {Leung},
  \citenamefont {Lu}, \citenamefont {Simon},\ and\ \citenamefont
  {Schuster}}]{Ma:2019aa}%
  \BibitemOpen
  \bibfield  {author} {\bibinfo {author} {\bibfnamefont {R.}~\bibnamefont
  {Ma}}, \bibinfo {author} {\bibfnamefont {B.}~\bibnamefont {Saxberg}},
  \bibinfo {author} {\bibfnamefont {C.}~\bibnamefont {Owens}}, \bibinfo
  {author} {\bibfnamefont {N.}~\bibnamefont {Leung}}, \bibinfo {author}
  {\bibfnamefont {Y.}~\bibnamefont {Lu}}, \bibinfo {author} {\bibfnamefont
  {J.}~\bibnamefont {Simon}},\ and\ \bibinfo {author} {\bibfnamefont {D.~I.}\
  \bibnamefont {Schuster}},\ }\href {https://doi.org/10.1038/s41586-019-0897-9}
  {\bibfield  {journal} {\bibinfo  {journal} {Nature}\ }\textbf {\bibinfo
  {volume} {566}},\ \bibinfo {pages} {51} (\bibinfo {year} {2019})}\BibitemShut
  {NoStop}%
\bibitem [{\citenamefont {Leger}\ \emph {et~al.}(2019)\citenamefont {Leger},
  \citenamefont {Puertas~Martinez}, \citenamefont {Bharadwaj}, \citenamefont
  {Dassonneville}, \citenamefont {Delaforce}, \citenamefont {Foroughi},
  \citenamefont {Milchakov}, \citenamefont {Planat}, \citenamefont {Buisson},
  \citenamefont {Naud}, \citenamefont {Hasch-Guichard}, \citenamefont
  {Florens}, \citenamefont {Snyman},\ and\ \citenamefont
  {Roch}}]{Leger:2019vv}%
  \BibitemOpen
  \bibfield  {author} {\bibinfo {author} {\bibfnamefont {S.}~\bibnamefont
  {Leger}}, \bibinfo {author} {\bibfnamefont {J.}~\bibnamefont
  {Puertas~Martinez}}, \bibinfo {author} {\bibfnamefont {K.}~\bibnamefont
  {Bharadwaj}}, \bibinfo {author} {\bibfnamefont {R.}~\bibnamefont
  {Dassonneville}}, \bibinfo {author} {\bibfnamefont {J.}~\bibnamefont
  {Delaforce}}, \bibinfo {author} {\bibfnamefont {F.}~\bibnamefont {Foroughi}},
  \bibinfo {author} {\bibfnamefont {V.}~\bibnamefont {Milchakov}}, \bibinfo
  {author} {\bibfnamefont {L.}~\bibnamefont {Planat}}, \bibinfo {author}
  {\bibfnamefont {O.}~\bibnamefont {Buisson}}, \bibinfo {author} {\bibfnamefont
  {C.}~\bibnamefont {Naud}}, \bibinfo {author} {\bibfnamefont {W.}~\bibnamefont
  {Hasch-Guichard}}, \bibinfo {author} {\bibfnamefont {S.}~\bibnamefont
  {Florens}}, \bibinfo {author} {\bibfnamefont {I.}~\bibnamefont {Snyman}},\
  and\ \bibinfo {author} {\bibfnamefont {N.}~\bibnamefont {Roch}},\ }\href
  {https://doi.org/10.1038/s41467-019-13199-x} {\bibfield  {journal} {\bibinfo
  {journal} {Nature Communications}\ }\textbf {\bibinfo {volume} {10}},\
  \bibinfo {pages} {5259} (\bibinfo {year} {2019})}\BibitemShut {NoStop}%
\bibitem [{\citenamefont {Carusotto}\ \emph {et~al.}(2020)\citenamefont
  {Carusotto}, \citenamefont {Houck}, \citenamefont {Koll{\'a}r}, \citenamefont
  {Roushan}, \citenamefont {Schuster},\ and\ \citenamefont
  {Simon}}]{Carusotto:2020ue}%
  \BibitemOpen
  \bibfield  {author} {\bibinfo {author} {\bibfnamefont {I.}~\bibnamefont
  {Carusotto}}, \bibinfo {author} {\bibfnamefont {A.~A.}\ \bibnamefont
  {Houck}}, \bibinfo {author} {\bibfnamefont {A.~J.}\ \bibnamefont
  {Koll{\'a}r}}, \bibinfo {author} {\bibfnamefont {P.}~\bibnamefont {Roushan}},
  \bibinfo {author} {\bibfnamefont {D.~I.}\ \bibnamefont {Schuster}},\ and\
  \bibinfo {author} {\bibfnamefont {J.}~\bibnamefont {Simon}},\ }\href
  {https://doi.org/10.1038/s41567-020-0815-y} {\bibfield  {journal} {\bibinfo
  {journal} {Nature Physics}\ }\textbf {\bibinfo {volume} {16}},\ \bibinfo
  {pages} {268} (\bibinfo {year} {2020})}\BibitemShut {NoStop}%
\bibitem [{\citenamefont {Feynman}\ \emph {et~al.}(2010)\citenamefont
  {Feynman}, \citenamefont {Leighton},\ and\ \citenamefont
  {Sands}}]{FeynmanLecture}%
  \BibitemOpen
  \bibfield  {author} {\bibinfo {author} {\bibfnamefont {R.~P.}\ \bibnamefont
  {Feynman}}, \bibinfo {author} {\bibfnamefont {R.~B.}\ \bibnamefont
  {Leighton}},\ and\ \bibinfo {author} {\bibfnamefont {M.}~\bibnamefont
  {Sands}},\ }\href {https://cds.cern.ch/record/1494701} {\emph {\bibinfo
  {title} {{The Feynman lectures on physics; New millennium ed.}}}}\ (\bibinfo
  {publisher} {Basic Books},\ \bibinfo {address} {New York, NY},\ \bibinfo
  {year} {2010})\BibitemShut {NoStop}%
\bibitem [{\citenamefont {Leggett}\ \emph {et~al.}(1987)\citenamefont
  {Leggett}, \citenamefont {Chakravarty}, \citenamefont {Dorsey}, \citenamefont
  {Fisher}, \citenamefont {Garg},\ and\ \citenamefont {Zwerger}}]{Leggett_RMP}%
  \BibitemOpen
  \bibfield  {author} {\bibinfo {author} {\bibfnamefont {A.~J.}\ \bibnamefont
  {Leggett}}, \bibinfo {author} {\bibfnamefont {S.}~\bibnamefont
  {Chakravarty}}, \bibinfo {author} {\bibfnamefont {A.~T.}\ \bibnamefont
  {Dorsey}}, \bibinfo {author} {\bibfnamefont {M.~P.~A.}\ \bibnamefont
  {Fisher}}, \bibinfo {author} {\bibfnamefont {A.}~\bibnamefont {Garg}},\ and\
  \bibinfo {author} {\bibfnamefont {W.}~\bibnamefont {Zwerger}},\ }\href@noop
  {} {\bibfield  {journal} {\bibinfo  {journal} {Reviews of Modern Physics}\
  }\textbf {\bibinfo {volume} {59}},\ \bibinfo {pages} {1} (\bibinfo {year}
  {1987})}\BibitemShut {NoStop}%
\bibitem [{\citenamefont {Weiss}(1992)}]{Weiss}%
  \BibitemOpen
  \bibfield  {author} {\bibinfo {author} {\bibfnamefont {U.}~\bibnamefont
  {Weiss}},\ }\href@noop {} {\emph {\bibinfo {title} {Quantum Dissipative
  Systems}}}\ (\bibinfo  {publisher} {World Scientific.},\ \bibinfo {year}
  {1992})\BibitemShut {NoStop}%
\bibitem [{\citenamefont {Puertas~Martinez}\ \emph {et~al.}(2019)\citenamefont
  {Puertas~Martinez}, \citenamefont {Leger}, \citenamefont {Gheeraert},
  \citenamefont {Dassonneville}, \citenamefont {Planat}, \citenamefont
  {Foroughi}, \citenamefont {Krupko}, \citenamefont {Buisson}, \citenamefont
  {Naud}, \citenamefont {Hasch-Guichard}, \citenamefont {Florens},
  \citenamefont {Snyman},\ and\ \citenamefont {Roch}}]{PuertasMartinez:2019gk}%
  \BibitemOpen
  \bibfield  {author} {\bibinfo {author} {\bibfnamefont {J.}~\bibnamefont
  {Puertas~Martinez}}, \bibinfo {author} {\bibfnamefont {S.}~\bibnamefont
  {Leger}}, \bibinfo {author} {\bibfnamefont {N.}~\bibnamefont {Gheeraert}},
  \bibinfo {author} {\bibfnamefont {R.}~\bibnamefont {Dassonneville}}, \bibinfo
  {author} {\bibfnamefont {L.}~\bibnamefont {Planat}}, \bibinfo {author}
  {\bibfnamefont {F.}~\bibnamefont {Foroughi}}, \bibinfo {author}
  {\bibfnamefont {Y.}~\bibnamefont {Krupko}}, \bibinfo {author} {\bibfnamefont
  {O.}~\bibnamefont {Buisson}}, \bibinfo {author} {\bibfnamefont
  {C.}~\bibnamefont {Naud}}, \bibinfo {author} {\bibfnamefont {W.}~\bibnamefont
  {Hasch-Guichard}}, \bibinfo {author} {\bibfnamefont {S.}~\bibnamefont
  {Florens}}, \bibinfo {author} {\bibfnamefont {I.}~\bibnamefont {Snyman}},\
  and\ \bibinfo {author} {\bibfnamefont {N.}~\bibnamefont {Roch}},\ }\href
  {https://doi.org/10.1038/s41534-018-0104-0} {\bibfield  {journal} {\bibinfo
  {journal} {npj Quantum Information}\ }\textbf {\bibinfo {volume} {5}},\
  \bibinfo {pages} {1829} (\bibinfo {year} {2019})}\BibitemShut {NoStop}%
\bibitem [{\citenamefont {Kuzmin}\ \emph {et~al.}(2019)\citenamefont {Kuzmin},
  \citenamefont {Mehta}, \citenamefont {Grabon}, \citenamefont {Mencia},\ and\
  \citenamefont {Manucharyan}}]{kuzmin_superstrong_2019}%
  \BibitemOpen
  \bibfield  {author} {\bibinfo {author} {\bibfnamefont {R.}~\bibnamefont
  {Kuzmin}}, \bibinfo {author} {\bibfnamefont {N.}~\bibnamefont {Mehta}},
  \bibinfo {author} {\bibfnamefont {N.}~\bibnamefont {Grabon}}, \bibinfo
  {author} {\bibfnamefont {R.}~\bibnamefont {Mencia}},\ and\ \bibinfo {author}
  {\bibfnamefont {V.~E.}\ \bibnamefont {Manucharyan}},\ }\href
  {https://doi.org/10.1038/s41534-019-0134-2} {\bibfield  {journal} {\bibinfo
  {journal} {npj Quantum Information}\ }\textbf {\bibinfo {volume} {5}},\
  \bibinfo {pages} {1} (\bibinfo {year} {2019})}\BibitemShut {NoStop}%
\bibitem [{\citenamefont {Weiss}\ \emph {et~al.}(2019)\citenamefont {Weiss},
  \citenamefont {Li}, \citenamefont {Ferguson},\ and\ \citenamefont
  {Koch}}]{Weiss:2019io}%
  \BibitemOpen
  \bibfield  {author} {\bibinfo {author} {\bibfnamefont {D.~K.}\ \bibnamefont
  {Weiss}}, \bibinfo {author} {\bibfnamefont {A.~C.~Y.}\ \bibnamefont {Li}},
  \bibinfo {author} {\bibfnamefont {D.~G.}\ \bibnamefont {Ferguson}},\ and\
  \bibinfo {author} {\bibfnamefont {J.}~\bibnamefont {Koch}},\ }\href
  {https://doi.org/10.1103/PhysRevB.100.224507} {\bibfield  {journal} {\bibinfo
   {journal} {Phys Rev B}\ }\textbf {\bibinfo {volume} {100}},\ \bibinfo
  {pages} {224507} (\bibinfo {year} {2019})}\BibitemShut {NoStop}%
\bibitem [{\citenamefont {Di~Paolo}\ \emph {et~al.}(2021)\citenamefont
  {Di~Paolo}, \citenamefont {Baker}, \citenamefont {Foley}, \citenamefont
  {S{\'e}n{\'e}chal},\ and\ \citenamefont {Blais}}]{DiPaolo:2019wl}%
  \BibitemOpen
  \bibfield  {author} {\bibinfo {author} {\bibfnamefont {A.}~\bibnamefont
  {Di~Paolo}}, \bibinfo {author} {\bibfnamefont {T.~E.}\ \bibnamefont {Baker}},
  \bibinfo {author} {\bibfnamefont {A.}~\bibnamefont {Foley}}, \bibinfo
  {author} {\bibfnamefont {D.}~\bibnamefont {S{\'e}n{\'e}chal}},\ and\ \bibinfo
  {author} {\bibfnamefont {A.}~\bibnamefont {Blais}},\ }\href
  {https://doi.org/10.1038/s41534-020-00352-4} {\bibfield  {journal} {\bibinfo
  {journal} {npj Quantum Inf.}\ }\textbf {\bibinfo {volume} {7}},\ \bibinfo
  {pages} {11} (\bibinfo {year} {2021})}\BibitemShut {NoStop}%
\bibitem [{\citenamefont {Grimsmo}\ \emph {et~al.}(2021)\citenamefont
  {Grimsmo}, \citenamefont {Royer}, \citenamefont {Kreikebaum}, \citenamefont
  {Ye}, \citenamefont {O'Brien}, \citenamefont {Siddiqi},\ and\ \citenamefont
  {Blais}}]{Grimsmo:2020va}%
  \BibitemOpen
  \bibfield  {author} {\bibinfo {author} {\bibfnamefont {A.~L.}\ \bibnamefont
  {Grimsmo}}, \bibinfo {author} {\bibfnamefont {B.}~\bibnamefont {Royer}},
  \bibinfo {author} {\bibfnamefont {J.~M.}\ \bibnamefont {Kreikebaum}},
  \bibinfo {author} {\bibfnamefont {Y.}~\bibnamefont {Ye}}, \bibinfo {author}
  {\bibfnamefont {K.}~\bibnamefont {O'Brien}}, \bibinfo {author} {\bibfnamefont
  {I.}~\bibnamefont {Siddiqi}},\ and\ \bibinfo {author} {\bibfnamefont
  {A.}~\bibnamefont {Blais}},\ }\href
  {https://doi.org/10.1103/PhysRevApplied.15.034074} {\bibfield  {journal}
  {\bibinfo  {journal} {Phys. Rev. Applied}\ }\textbf {\bibinfo {volume}
  {15}},\ \bibinfo {pages} {034074} (\bibinfo {year} {2021})}\BibitemShut
  {NoStop}%
\bibitem [{\citenamefont {Kjaergaard}\ \emph {et~al.}(2020)\citenamefont
  {Kjaergaard}, \citenamefont {Schwartz}, \citenamefont {Braumüller},
  \citenamefont {Krantz}, \citenamefont {Wang}, \citenamefont {Gustavsson},\
  and\ \citenamefont {Oliver}}]{ReviewSuperconductingSimulators}%
  \BibitemOpen
  \bibfield  {author} {\bibinfo {author} {\bibfnamefont {M.}~\bibnamefont
  {Kjaergaard}}, \bibinfo {author} {\bibfnamefont {M.~E.}\ \bibnamefont
  {Schwartz}}, \bibinfo {author} {\bibfnamefont {J.}~\bibnamefont
  {Braumüller}}, \bibinfo {author} {\bibfnamefont {P.}~\bibnamefont {Krantz}},
  \bibinfo {author} {\bibfnamefont {J.~I.-J.}\ \bibnamefont {Wang}}, \bibinfo
  {author} {\bibfnamefont {S.}~\bibnamefont {Gustavsson}},\ and\ \bibinfo
  {author} {\bibfnamefont {W.~D.}\ \bibnamefont {Oliver}},\ }\href
  {https://doi.org/10.1146/annurev-conmatphys-031119-050605} {\bibfield
  {journal} {\bibinfo  {journal} {Annual Review of Condensed Matter Physics}\
  }\textbf {\bibinfo {volume} {11}},\ \bibinfo {pages} {369} (\bibinfo {year}
  {2020})},\ \Eprint
  {https://arxiv.org/abs/https://doi.org/10.1146/annurev-conmatphys-031119-050605}
  {https://doi.org/10.1146/annurev-conmatphys-031119-050605} \BibitemShut
  {NoStop}%
\bibitem [{\citenamefont {Cedraschi}\ \emph {et~al.}(2000)\citenamefont
  {Cedraschi}, \citenamefont {Ponomarenko},\ and\ \citenamefont
  {B\"uttiker}}]{Cedraschi_Ring}%
  \BibitemOpen
  \bibfield  {author} {\bibinfo {author} {\bibfnamefont {P.}~\bibnamefont
  {Cedraschi}}, \bibinfo {author} {\bibfnamefont {V.~V.}\ \bibnamefont
  {Ponomarenko}},\ and\ \bibinfo {author} {\bibfnamefont {M.}~\bibnamefont
  {B\"uttiker}},\ }\href {https://doi.org/10.1103/PhysRevLett.84.346}
  {\bibfield  {journal} {\bibinfo  {journal} {Phys. Rev. Lett.}\ }\textbf
  {\bibinfo {volume} {84}},\ \bibinfo {pages} {346} (\bibinfo {year}
  {2000})}\BibitemShut {NoStop}%
\bibitem [{\citenamefont {Furusaki}\ and\ \citenamefont
  {Matveev}(2002)}]{Matveev_Resonant}%
  \BibitemOpen
  \bibfield  {author} {\bibinfo {author} {\bibfnamefont {A.}~\bibnamefont
  {Furusaki}}\ and\ \bibinfo {author} {\bibfnamefont {K.~A.}\ \bibnamefont
  {Matveev}},\ }\href {https://doi.org/10.1103/PhysRevLett.88.226404}
  {\bibfield  {journal} {\bibinfo  {journal} {Phys. Rev. Lett.}\ }\textbf
  {\bibinfo {volume} {88}},\ \bibinfo {pages} {226404} (\bibinfo {year}
  {2002})}\BibitemShut {NoStop}%
\bibitem [{\citenamefont {Le~Hur}\ and\ \citenamefont
  {Li}(2005)}]{LeHur_Resonant}%
  \BibitemOpen
  \bibfield  {author} {\bibinfo {author} {\bibfnamefont {K.}~\bibnamefont
  {Le~Hur}}\ and\ \bibinfo {author} {\bibfnamefont {M.-R.}\ \bibnamefont
  {Li}},\ }\href {https://doi.org/10.1103/PhysRevB.72.073305} {\bibfield
  {journal} {\bibinfo  {journal} {Phys. Rev. B}\ }\textbf {\bibinfo {volume}
  {72}},\ \bibinfo {pages} {073305} (\bibinfo {year} {2005})}\BibitemShut
  {NoStop}%
\bibitem [{\citenamefont {Le~Hur}(2004)}]{LeHur_Box}%
  \BibitemOpen
  \bibfield  {author} {\bibinfo {author} {\bibfnamefont {K.}~\bibnamefont
  {Le~Hur}},\ }\href {https://doi.org/10.1103/PhysRevLett.92.196804} {\bibfield
   {journal} {\bibinfo  {journal} {Phys. Rev. Lett.}\ }\textbf {\bibinfo
  {volume} {92}},\ \bibinfo {pages} {196804} (\bibinfo {year}
  {2004})}\BibitemShut {NoStop}%
\bibitem [{\citenamefont {Li}\ \emph {et~al.}(2005)\citenamefont {Li},
  \citenamefont {Le~Hur},\ and\ \citenamefont {Hofstetter}}]{LeHur_Caldeira}%
  \BibitemOpen
  \bibfield  {author} {\bibinfo {author} {\bibfnamefont {M.-R.}\ \bibnamefont
  {Li}}, \bibinfo {author} {\bibfnamefont {K.}~\bibnamefont {Le~Hur}},\ and\
  \bibinfo {author} {\bibfnamefont {W.}~\bibnamefont {Hofstetter}},\ }\href
  {https://doi.org/10.1103/PhysRevLett.95.086406} {\bibfield  {journal}
  {\bibinfo  {journal} {Phys. Rev. Lett.}\ }\textbf {\bibinfo {volume} {95}},\
  \bibinfo {pages} {086406} (\bibinfo {year} {2005})}\BibitemShut {NoStop}%
\bibitem [{\citenamefont {Recati}\ \emph {et~al.}(2005)\citenamefont {Recati},
  \citenamefont {Fedichev}, \citenamefont {Zwerger}, \citenamefont {von
  Delft},\ and\ \citenamefont {Zoller}}]{Recati_Atomic}%
  \BibitemOpen
  \bibfield  {author} {\bibinfo {author} {\bibfnamefont {A.}~\bibnamefont
  {Recati}}, \bibinfo {author} {\bibfnamefont {P.~O.}\ \bibnamefont
  {Fedichev}}, \bibinfo {author} {\bibfnamefont {W.}~\bibnamefont {Zwerger}},
  \bibinfo {author} {\bibfnamefont {J.}~\bibnamefont {von Delft}},\ and\
  \bibinfo {author} {\bibfnamefont {P.}~\bibnamefont {Zoller}},\ }\href
  {https://doi.org/10.1103/PhysRevLett.94.040404} {\bibfield  {journal}
  {\bibinfo  {journal} {Phys. Rev. Lett.}\ }\textbf {\bibinfo {volume} {94}},\
  \bibinfo {pages} {040404} (\bibinfo {year} {2005})}\BibitemShut {NoStop}%
\bibitem [{\citenamefont {Lescanne}\ \emph {et~al.}(2019)\citenamefont
  {Lescanne}, \citenamefont {Verney}, \citenamefont {Ficheux}, \citenamefont
  {Devoret}, \citenamefont {Huard}, \citenamefont {Mirrahimi},\ and\
  \citenamefont {Leghtas}}]{Verney}%
  \BibitemOpen
  \bibfield  {author} {\bibinfo {author} {\bibfnamefont {R.}~\bibnamefont
  {Lescanne}}, \bibinfo {author} {\bibfnamefont {L.}~\bibnamefont {Verney}},
  \bibinfo {author} {\bibfnamefont {Q.}~\bibnamefont {Ficheux}}, \bibinfo
  {author} {\bibfnamefont {M.~H.}\ \bibnamefont {Devoret}}, \bibinfo {author}
  {\bibfnamefont {B.}~\bibnamefont {Huard}}, \bibinfo {author} {\bibfnamefont
  {M.}~\bibnamefont {Mirrahimi}},\ and\ \bibinfo {author} {\bibfnamefont
  {Z.}~\bibnamefont {Leghtas}},\ }\href
  {https://doi.org/10.1103/PhysRevApplied.11.014030} {\bibfield  {journal}
  {\bibinfo  {journal} {Phys. Rev. Applied}\ }\textbf {\bibinfo {volume}
  {11}},\ \bibinfo {pages} {014030} (\bibinfo {year} {2019})}\BibitemShut
  {NoStop}%
\bibitem [{\citenamefont {Nigg}\ \emph {et~al.}(2012)\citenamefont {Nigg},
  \citenamefont {Paik}, \citenamefont {Vlastakis}, \citenamefont {Kirchmair},
  \citenamefont {Shankar}, \citenamefont {Frunzio}, \citenamefont {Devoret},
  \citenamefont {Schoelkopf},\ and\ \citenamefont {Girvin}}]{Anonymous:2012ek}%
  \BibitemOpen
  \bibfield  {author} {\bibinfo {author} {\bibfnamefont {S.~E.}\ \bibnamefont
  {Nigg}}, \bibinfo {author} {\bibfnamefont {H.}~\bibnamefont {Paik}}, \bibinfo
  {author} {\bibfnamefont {B.}~\bibnamefont {Vlastakis}}, \bibinfo {author}
  {\bibfnamefont {G.}~\bibnamefont {Kirchmair}}, \bibinfo {author}
  {\bibfnamefont {S.}~\bibnamefont {Shankar}}, \bibinfo {author} {\bibfnamefont
  {L.}~\bibnamefont {Frunzio}}, \bibinfo {author} {\bibfnamefont {M.~H.}\
  \bibnamefont {Devoret}}, \bibinfo {author} {\bibfnamefont {R.~J.}\
  \bibnamefont {Schoelkopf}},\ and\ \bibinfo {author} {\bibfnamefont {S.~M.}\
  \bibnamefont {Girvin}},\ }\href
  {https://doi.org/10.1103/physrevlett.108.240502} {\bibfield  {journal}
  {\bibinfo  {journal} {Physical Review Letters}\ }\textbf {\bibinfo {volume}
  {108}},\ \bibinfo {pages} {240502} (\bibinfo {year} {2012})}\BibitemShut
  {NoStop}%
\bibitem [{\citenamefont {Garcia-Ripoll}\ \emph {et~al.}(2015)\citenamefont
  {Garcia-Ripoll}, \citenamefont {Peropadre},\ and\ \citenamefont
  {De~Liberato}}]{GarciaRipoll:2015ba}%
  \BibitemOpen
  \bibfield  {author} {\bibinfo {author} {\bibfnamefont {J.~J.}\ \bibnamefont
  {Garcia-Ripoll}}, \bibinfo {author} {\bibfnamefont {B.}~\bibnamefont
  {Peropadre}},\ and\ \bibinfo {author} {\bibfnamefont {S.}~\bibnamefont
  {De~Liberato}},\ }\href {https://doi.org/10.1038/srep16055} {\bibfield
  {journal} {\bibinfo  {journal} {Scientific Reports}\ }\textbf {\bibinfo
  {volume} {5}},\ \bibinfo {pages} {16055} (\bibinfo {year}
  {2015})}\BibitemShut {NoStop}%
\bibitem [{\citenamefont {Malekakhlagh}\ \emph {et~al.}(2017)\citenamefont
  {Malekakhlagh}, \citenamefont {Petrescu},\ and\ \citenamefont
  {T{\"u}reci}}]{malekakhlagh_cutoff-free_2017}%
  \BibitemOpen
  \bibfield  {author} {\bibinfo {author} {\bibfnamefont {M.}~\bibnamefont
  {Malekakhlagh}}, \bibinfo {author} {\bibfnamefont {A.}~\bibnamefont
  {Petrescu}},\ and\ \bibinfo {author} {\bibfnamefont {H.~E.}\ \bibnamefont
  {T{\"u}reci}},\ }\href {https://doi.org/10.1103/PhysRevLett.119.073601}
  {\bibfield  {journal} {\bibinfo  {journal} {Physical Review Letters}\
  }\textbf {\bibinfo {volume} {119}},\ \bibinfo {pages} {073601} (\bibinfo
  {year} {2017})}\BibitemShut {NoStop}%
\bibitem [{\citenamefont {Parra-Rodriguez}\ \emph {et~al.}(2018)\citenamefont
  {Parra-Rodriguez}, \citenamefont {Rico}, \citenamefont {Solano},\ and\
  \citenamefont {Egusquiza}}]{ParraRodriguez:2018da}%
  \BibitemOpen
  \bibfield  {author} {\bibinfo {author} {\bibfnamefont {A.}~\bibnamefont
  {Parra-Rodriguez}}, \bibinfo {author} {\bibfnamefont {E.}~\bibnamefont
  {Rico}}, \bibinfo {author} {\bibfnamefont {E.}~\bibnamefont {Solano}},\ and\
  \bibinfo {author} {\bibfnamefont {I.~L.}\ \bibnamefont {Egusquiza}},\ }\href
  {https://doi.org/10.1088/2058-9565/aab1ba} {\bibfield  {journal} {\bibinfo
  {journal} {Quantum Science and Technology}\ }\textbf {\bibinfo {volume}
  {3}},\ \bibinfo {pages} {024012} (\bibinfo {year} {2018})}\BibitemShut
  {NoStop}%
\bibitem [{\citenamefont {Forn-D{\'\i}az}\ \emph {et~al.}(2019)\citenamefont
  {Forn-D{\'\i}az}, \citenamefont {Lamata}, \citenamefont {Rico}, \citenamefont
  {Kono},\ and\ \citenamefont {Solano}}]{FornDiaz:2019br}%
  \BibitemOpen
  \bibfield  {author} {\bibinfo {author} {\bibfnamefont {P.}~\bibnamefont
  {Forn-D{\'\i}az}}, \bibinfo {author} {\bibfnamefont {L.}~\bibnamefont
  {Lamata}}, \bibinfo {author} {\bibfnamefont {E.}~\bibnamefont {Rico}},
  \bibinfo {author} {\bibfnamefont {J.}~\bibnamefont {Kono}},\ and\ \bibinfo
  {author} {\bibfnamefont {E.}~\bibnamefont {Solano}},\ }\href
  {https://doi.org/10.1103/RevModPhys.91.025005} {\bibfield  {journal}
  {\bibinfo  {journal} {Reviews of Modern Physics}\ }\textbf {\bibinfo {volume}
  {91}},\ \bibinfo {pages} {025005} (\bibinfo {year} {2019})}\BibitemShut
  {NoStop}%
\bibitem [{\citenamefont {Kockum}\ \emph {et~al.}(2019)\citenamefont {Kockum},
  \citenamefont {Miranowicz}, \citenamefont {De~Liberato}, \citenamefont
  {Savasta},\ and\ \citenamefont {Nori}}]{Kockum:2019ky}%
  \BibitemOpen
  \bibfield  {author} {\bibinfo {author} {\bibfnamefont {A.~F.}\ \bibnamefont
  {Kockum}}, \bibinfo {author} {\bibfnamefont {A.}~\bibnamefont {Miranowicz}},
  \bibinfo {author} {\bibfnamefont {S.}~\bibnamefont {De~Liberato}}, \bibinfo
  {author} {\bibfnamefont {S.}~\bibnamefont {Savasta}},\ and\ \bibinfo {author}
  {\bibfnamefont {F.}~\bibnamefont {Nori}},\ }\href
  {https://doi.org/10.1038/s42254-018-0006-2} {\bibfield  {journal} {\bibinfo
  {journal} {Nature Reviews Physics}\ }\textbf {\bibinfo {volume} {1}},\
  \bibinfo {pages} {19} (\bibinfo {year} {2019})}\BibitemShut {NoStop}%
\bibitem [{\citenamefont {Le~Boit\'e}(2020)}]{LeBoite}%
  \BibitemOpen
  \bibfield  {author} {\bibinfo {author} {\bibfnamefont {A.}~\bibnamefont
  {Le~Boit\'e}},\ }\href {https://doi.org/10.1002/qute.201900140} {\bibfield
  {journal} {\bibinfo  {journal} {Advanced Quantum Technologies}\ }\textbf
  {\bibinfo {volume} {3}},\ \bibinfo {pages} {1900140} (\bibinfo {year}
  {2020})}\BibitemShut {NoStop}%
\bibitem [{\citenamefont {Koch}\ \emph {et~al.}(2007)\citenamefont {Koch},
  \citenamefont {Yu}, \citenamefont {Gambetta}, \citenamefont {Houck},
  \citenamefont {Schuster}, \citenamefont {Majer}, \citenamefont {Blais},
  \citenamefont {Devoret}, \citenamefont {Girvin},\ and\ \citenamefont
  {Schoelkopf}}]{Koch_Transmon}%
  \BibitemOpen
  \bibfield  {author} {\bibinfo {author} {\bibfnamefont {J.}~\bibnamefont
  {Koch}}, \bibinfo {author} {\bibfnamefont {T.~M.}\ \bibnamefont {Yu}},
  \bibinfo {author} {\bibfnamefont {J.}~\bibnamefont {Gambetta}}, \bibinfo
  {author} {\bibfnamefont {A.~A.}\ \bibnamefont {Houck}}, \bibinfo {author}
  {\bibfnamefont {D.~I.}\ \bibnamefont {Schuster}}, \bibinfo {author}
  {\bibfnamefont {J.}~\bibnamefont {Majer}}, \bibinfo {author} {\bibfnamefont
  {A.}~\bibnamefont {Blais}}, \bibinfo {author} {\bibfnamefont {M.~H.}\
  \bibnamefont {Devoret}}, \bibinfo {author} {\bibfnamefont {S.~M.}\
  \bibnamefont {Girvin}},\ and\ \bibinfo {author} {\bibfnamefont {R.~J.}\
  \bibnamefont {Schoelkopf}},\ }\href
  {https://doi.org/10.1103/PhysRevA.76.042319} {\bibfield  {journal} {\bibinfo
  {journal} {Phys. Rev. A}\ }\textbf {\bibinfo {volume} {76}},\ \bibinfo
  {pages} {042319} (\bibinfo {year} {2007})}\BibitemShut {NoStop}%
\bibitem [{\citenamefont {Glazman}\ and\ \citenamefont
  {Larkin}(1997)}]{GlazmanLarkin}%
  \BibitemOpen
  \bibfield  {author} {\bibinfo {author} {\bibfnamefont {L.~I.}\ \bibnamefont
  {Glazman}}\ and\ \bibinfo {author} {\bibfnamefont {A.~I.}\ \bibnamefont
  {Larkin}},\ }\href {https://doi.org/10.1103/PhysRevLett.79.3736} {\bibfield
  {journal} {\bibinfo  {journal} {Phys. Rev. Lett.}\ }\textbf {\bibinfo
  {volume} {79}},\ \bibinfo {pages} {3736} (\bibinfo {year}
  {1997})}\BibitemShut {NoStop}%
\bibitem [{\citenamefont {Basko}\ \emph {et~al.}(2020)\citenamefont {Basko},
  \citenamefont {Pfeiffer}, \citenamefont {Adamus}, \citenamefont {Holzmann},\
  and\ \citenamefont {Hekking}}]{Basko_SIT}%
  \BibitemOpen
  \bibfield  {author} {\bibinfo {author} {\bibfnamefont {D.~M.}\ \bibnamefont
  {Basko}}, \bibinfo {author} {\bibfnamefont {F.}~\bibnamefont {Pfeiffer}},
  \bibinfo {author} {\bibfnamefont {P.}~\bibnamefont {Adamus}}, \bibinfo
  {author} {\bibfnamefont {M.}~\bibnamefont {Holzmann}},\ and\ \bibinfo
  {author} {\bibfnamefont {F.~W.~J.}\ \bibnamefont {Hekking}},\ }\href
  {https://doi.org/10.1103/PhysRevB.101.024518} {\bibfield  {journal} {\bibinfo
   {journal} {Phys. Rev. B}\ }\textbf {\bibinfo {volume} {101}},\ \bibinfo
  {pages} {024518} (\bibinfo {year} {2020})}\BibitemShut {NoStop}%
\bibitem [{\citenamefont {Roy}\ \emph {et~al.}(2021)\citenamefont {Roy},
  \citenamefont {Schuricht}, \citenamefont {Hauschild}, \citenamefont
  {Pollmann},\ and\ \citenamefont {Saleur}}]{Roy_SineGordonSimulation}%
  \BibitemOpen
  \bibfield  {author} {\bibinfo {author} {\bibfnamefont {A.}~\bibnamefont
  {Roy}}, \bibinfo {author} {\bibfnamefont {D.}~\bibnamefont {Schuricht}},
  \bibinfo {author} {\bibfnamefont {J.}~\bibnamefont {Hauschild}}, \bibinfo
  {author} {\bibfnamefont {F.}~\bibnamefont {Pollmann}},\ and\ \bibinfo
  {author} {\bibfnamefont {H.}~\bibnamefont {Saleur}},\ }\href
  {https://doi.org/https://doi.org/10.1016/j.nuclphysb.2021.115445} {\bibfield
  {journal} {\bibinfo  {journal} {Nuclear Physics B}\ }\textbf {\bibinfo
  {volume} {968}},\ \bibinfo {pages} {115445} (\bibinfo {year}
  {2021})}\BibitemShut {NoStop}%
\bibitem [{InP()}]{InPrep}%
  \BibitemOpen
  \href@noop {} {}\bibinfo {note} {K. Kaur {\it et al.}, in
  preparation.}\BibitemShut {Stop}%
\bibitem [{\citenamefont {Le~Hur}(2012)}]{LeHur_Kondo_2012}%
  \BibitemOpen
  \bibfield  {author} {\bibinfo {author} {\bibfnamefont {K.}~\bibnamefont
  {Le~Hur}},\ }\href@noop {} {\bibfield  {journal} {\bibinfo  {journal} {Phys.
  Rev. B}\ }\textbf {\bibinfo {volume} {85}},\ \bibinfo {pages} {140506(R)}
  (\bibinfo {year} {2012})}\BibitemShut {NoStop}%
\bibitem [{\citenamefont {Goldstein}\ \emph {et~al.}(2013)\citenamefont
  {Goldstein}, \citenamefont {Devoret}, \citenamefont {Houzet},\ and\
  \citenamefont {Glazman}}]{goldstein_inelastic_2013}%
  \BibitemOpen
  \bibfield  {author} {\bibinfo {author} {\bibfnamefont {M.}~\bibnamefont
  {Goldstein}}, \bibinfo {author} {\bibfnamefont {M.~H.}\ \bibnamefont
  {Devoret}}, \bibinfo {author} {\bibfnamefont {M.}~\bibnamefont {Houzet}},\
  and\ \bibinfo {author} {\bibfnamefont {L.~I.}\ \bibnamefont {Glazman}},\
  }\href@noop {} {\bibfield  {journal} {\bibinfo  {journal} {Physical Review
  Letters}\ }\textbf {\bibinfo {volume} {110}},\ \bibinfo {pages} {017002}
  (\bibinfo {year} {2013})}\BibitemShut {NoStop}%
\bibitem [{\citenamefont {Peropadre}\ \emph {et~al.}(2013)\citenamefont
  {Peropadre}, \citenamefont {Zueco}, \citenamefont {Porras},\ and\
  \citenamefont {Garcia-Ripoll}}]{peropadre_2013}%
  \BibitemOpen
  \bibfield  {author} {\bibinfo {author} {\bibfnamefont {B.}~\bibnamefont
  {Peropadre}}, \bibinfo {author} {\bibfnamefont {D.}~\bibnamefont {Zueco}},
  \bibinfo {author} {\bibfnamefont {D.}~\bibnamefont {Porras}},\ and\ \bibinfo
  {author} {\bibfnamefont {J.~J.}\ \bibnamefont {Garcia-Ripoll}},\ }\href@noop
  {} {\bibfield  {journal} {\bibinfo  {journal} {Phys. Rev. Lett.}\ }\textbf
  {\bibinfo {volume} {111}},\ \bibinfo {pages} {243602} (\bibinfo {year}
  {2013})}\BibitemShut {NoStop}%
\bibitem [{\citenamefont {Snyman}\ and\ \citenamefont
  {Florens}(2015)}]{snyman_robust_2015}%
  \BibitemOpen
  \bibfield  {author} {\bibinfo {author} {\bibfnamefont {I.}~\bibnamefont
  {Snyman}}\ and\ \bibinfo {author} {\bibfnamefont {S.}~\bibnamefont
  {Florens}},\ }\href@noop {} {\bibfield  {journal} {\bibinfo  {journal}
  {Physical Review B}\ }\textbf {\bibinfo {volume} {92}},\ \bibinfo {pages}
  {085131} (\bibinfo {year} {2015})}\BibitemShut {NoStop}%
\bibitem [{\citenamefont {Gheeraert}\ \emph {et~al.}(2018)\citenamefont
  {Gheeraert}, \citenamefont {Zhang}, \citenamefont {S{\'e}pulcre},
  \citenamefont {Bera}, \citenamefont {Roch}, \citenamefont {Baranger},\ and\
  \citenamefont {Florens}}]{Gheeraert:2018bv}%
  \BibitemOpen
  \bibfield  {author} {\bibinfo {author} {\bibfnamefont {N.}~\bibnamefont
  {Gheeraert}}, \bibinfo {author} {\bibfnamefont {X.~H.~H.}\ \bibnamefont
  {Zhang}}, \bibinfo {author} {\bibfnamefont {T.}~\bibnamefont {S{\'e}pulcre}},
  \bibinfo {author} {\bibfnamefont {S.}~\bibnamefont {Bera}}, \bibinfo {author}
  {\bibfnamefont {N.}~\bibnamefont {Roch}}, \bibinfo {author} {\bibfnamefont
  {H.~U.}\ \bibnamefont {Baranger}},\ and\ \bibinfo {author} {\bibfnamefont
  {S.}~\bibnamefont {Florens}},\ }\href@noop {} {\bibfield  {journal} {\bibinfo
   {journal} {Physical Review A}\ }\textbf {\bibinfo {volume} {98}},\ \bibinfo
  {pages} {043816} (\bibinfo {year} {2018})}\BibitemShut {NoStop}%
\bibitem [{\citenamefont {Magazz\`u}\ \emph {et~al.}(2018)\citenamefont
  {Magazz\`u}, \citenamefont {Forn-D\'iaz}, \citenamefont {Belyansky},
  \citenamefont {Orgiazzi}, \citenamefont {Yurtalan}, \citenamefont {Otto},
  \citenamefont {Lupascu}, \citenamefont {Wilson},\ and\ \citenamefont
  {Grifoni}}]{Magazzu}%
  \BibitemOpen
  \bibfield  {author} {\bibinfo {author} {\bibfnamefont {L.}~\bibnamefont
  {Magazz\`u}}, \bibinfo {author} {\bibfnamefont {P.}~\bibnamefont
  {Forn-D\'iaz}}, \bibinfo {author} {\bibfnamefont {R.}~\bibnamefont
  {Belyansky}}, \bibinfo {author} {\bibfnamefont {J.-L.}\ \bibnamefont
  {Orgiazzi}}, \bibinfo {author} {\bibfnamefont {M.~A.}\ \bibnamefont
  {Yurtalan}}, \bibinfo {author} {\bibfnamefont {M.~R.}\ \bibnamefont {Otto}},
  \bibinfo {author} {\bibfnamefont {A.}~\bibnamefont {Lupascu}}, \bibinfo
  {author} {\bibfnamefont {C.~M.}\ \bibnamefont {Wilson}},\ and\ \bibinfo
  {author} {\bibfnamefont {M.}~\bibnamefont {Grifoni}},\ }\href
  {https://doi.org/10.1038/s41467-018-03626-w} {\bibfield  {journal} {\bibinfo
  {journal} {Nature Communications}\ }\textbf {\bibinfo {volume} {9}},\
  \bibinfo {pages} {1403} (\bibinfo {year} {2018})}\BibitemShut {NoStop}%
\bibitem [{\citenamefont {Vool}\ and\ \citenamefont
  {Devoret}(2017)}]{vool_introduction_2017}%
  \BibitemOpen
  \bibfield  {author} {\bibinfo {author} {\bibfnamefont {U.}~\bibnamefont
  {Vool}}\ and\ \bibinfo {author} {\bibfnamefont {M.~H.}\ \bibnamefont
  {Devoret}},\ }\href {https://doi.org/10.1002/cta.2359} {\bibfield  {journal}
  {\bibinfo  {journal} {International Journal of Circuit Theory and
  Applications}\ }\textbf {\bibinfo {volume} {45}},\ \bibinfo {pages} {897}
  (\bibinfo {year} {2017})}\BibitemShut {NoStop}%
\bibitem [{Sup()}]{SupInfo}%
  \BibitemOpen
  \href@noop {} {}\bibinfo {note} {See supplementary informations for
  details.}\BibitemShut {Stop}%
\bibitem [{\citenamefont {Zwerger}\ \emph {et~al.}(1986)\citenamefont
  {Zwerger}, \citenamefont {Dorsey},\ and\ \citenamefont
  {Fisher}}]{Zwerger_Periodicity}%
  \BibitemOpen
  \bibfield  {author} {\bibinfo {author} {\bibfnamefont {W.}~\bibnamefont
  {Zwerger}}, \bibinfo {author} {\bibfnamefont {A.~T.}\ \bibnamefont
  {Dorsey}},\ and\ \bibinfo {author} {\bibfnamefont {M.~P.~A.}\ \bibnamefont
  {Fisher}},\ }\href {https://doi.org/10.1103/PhysRevB.34.6518} {\bibfield
  {journal} {\bibinfo  {journal} {Phys. Rev. B}\ }\textbf {\bibinfo {volume}
  {34}},\ \bibinfo {pages} {6518} (\bibinfo {year} {1986})}\BibitemShut
  {NoStop}%
\bibitem [{\citenamefont {Sch{\"o}n}\ and\ \citenamefont
  {Zaikin}(1990)}]{SchoenZaikin}%
  \BibitemOpen
  \bibfield  {author} {\bibinfo {author} {\bibfnamefont {G.}~\bibnamefont
  {Sch{\"o}n}}\ and\ \bibinfo {author} {\bibfnamefont {A.~D.}\ \bibnamefont
  {Zaikin}},\ }\href@noop {} {\bibfield  {journal} {\bibinfo  {journal}
  {Physics Reports}\ }\textbf {\bibinfo {volume} {198}},\ \bibinfo {pages}
  {237} (\bibinfo {year} {1990})}\BibitemShut {NoStop}%
\bibitem [{\citenamefont {Bosman}\ \emph {et~al.}(2017)\citenamefont {Bosman},
  \citenamefont {Gely}, \citenamefont {Singh}, \citenamefont {Bothner},
  \citenamefont {Castellanos-Gomez},\ and\ \citenamefont
  {Steele}}]{Bosman:2017el}%
  \BibitemOpen
  \bibfield  {author} {\bibinfo {author} {\bibfnamefont {S.~J.}\ \bibnamefont
  {Bosman}}, \bibinfo {author} {\bibfnamefont {M.~F.}\ \bibnamefont {Gely}},
  \bibinfo {author} {\bibfnamefont {V.}~\bibnamefont {Singh}}, \bibinfo
  {author} {\bibfnamefont {D.}~\bibnamefont {Bothner}}, \bibinfo {author}
  {\bibfnamefont {A.}~\bibnamefont {Castellanos-Gomez}},\ and\ \bibinfo
  {author} {\bibfnamefont {G.~A.}\ \bibnamefont {Steele}},\ }\href
  {https://doi.org/10.1103/PhysRevB.95.224515} {\bibfield  {journal} {\bibinfo
  {journal} {Phys Rev B}\ }\textbf {\bibinfo {volume} {95}},\ \bibinfo {pages}
  {224515} (\bibinfo {year} {2017})}\BibitemShut {NoStop}%
\bibitem [{\citenamefont {Jaako}\ \emph {et~al.}(2016)\citenamefont {Jaako},
  \citenamefont {Xiang}, \citenamefont {Garcia-Ripoll},\ and\ \citenamefont
  {Rabl}}]{Jaako:2016io}%
  \BibitemOpen
  \bibfield  {author} {\bibinfo {author} {\bibfnamefont {T.}~\bibnamefont
  {Jaako}}, \bibinfo {author} {\bibfnamefont {Z.-L.}\ \bibnamefont {Xiang}},
  \bibinfo {author} {\bibfnamefont {J.~J.}\ \bibnamefont {Garcia-Ripoll}},\
  and\ \bibinfo {author} {\bibfnamefont {P.}~\bibnamefont {Rabl}},\ }\href
  {https://doi.org/10.1103/PhysRevA.94.033850} {\bibfield  {journal} {\bibinfo
  {journal} {Physical Review A}\ }\textbf {\bibinfo {volume} {94}},\ \bibinfo
  {pages} {033850} (\bibinfo {year} {2016})}\BibitemShut {NoStop}%
\bibitem [{\citenamefont {Nataf}\ and\ \citenamefont
  {Ciuti}(2010)}]{DickeNoGo}%
  \BibitemOpen
  \bibfield  {author} {\bibinfo {author} {\bibfnamefont {P.}~\bibnamefont
  {Nataf}}\ and\ \bibinfo {author} {\bibfnamefont {C.}~\bibnamefont {Ciuti}},\
  }\href@noop {} {\bibfield  {journal} {\bibinfo  {journal} {Nat. Commun.}\
  }\textbf {\bibinfo {volume} {1}},\ \bibinfo {pages} {72} (\bibinfo {year}
  {2010})}\BibitemShut {NoStop}%
\bibitem [{\citenamefont {Bulla}\ \emph {et~al.}(2008)\citenamefont {Bulla},
  \citenamefont {Costi},\ and\ \citenamefont {Pruschke}}]{NRG-RMP08}%
  \BibitemOpen
  \bibfield  {author} {\bibinfo {author} {\bibfnamefont {R.}~\bibnamefont
  {Bulla}}, \bibinfo {author} {\bibfnamefont {T.~A.}\ \bibnamefont {Costi}},\
  and\ \bibinfo {author} {\bibfnamefont {T.}~\bibnamefont {Pruschke}},\ }\href
  {https://doi.org/10.1103/RevModPhys.80.395} {\bibfield  {journal} {\bibinfo
  {journal} {Rev. Mod. Phys.}\ }\textbf {\bibinfo {volume} {80}},\ \bibinfo
  {pages} {395} (\bibinfo {year} {2008})}\BibitemShut {NoStop}%
\bibitem [{\citenamefont {Bulla}\ \emph {et~al.}(2003)\citenamefont {Bulla},
  \citenamefont {Tong},\ and\ \citenamefont {Vojta}}]{Tong}%
  \BibitemOpen
  \bibfield  {author} {\bibinfo {author} {\bibfnamefont {R.}~\bibnamefont
  {Bulla}}, \bibinfo {author} {\bibfnamefont {N.-H.}\ \bibnamefont {Tong}},\
  and\ \bibinfo {author} {\bibfnamefont {M.}~\bibnamefont {Vojta}},\ }\href
  {https://doi.org/10.1103/PhysRevLett.91.170601} {\bibfield  {journal}
  {\bibinfo  {journal} {Phys. Rev. Lett.}\ }\textbf {\bibinfo {volume} {91}},\
  \bibinfo {pages} {170601} (\bibinfo {year} {2003})}\BibitemShut {NoStop}%
\bibitem [{\citenamefont {Silbey}\ and\ \citenamefont {Harris}(1984)}]{Silbey}%
  \BibitemOpen
  \bibfield  {author} {\bibinfo {author} {\bibfnamefont {R.}~\bibnamefont
  {Silbey}}\ and\ \bibinfo {author} {\bibfnamefont {R.~A.}\ \bibnamefont
  {Harris}},\ }\href {https://doi.org/http://dx.doi.org/10.1063/1.447055}
  {\bibfield  {journal} {\bibinfo  {journal} {J. Chem. Phys.}\ }\textbf
  {\bibinfo {volume} {80}},\ \bibinfo {pages} {2615} (\bibinfo {year}
  {1984})}\BibitemShut {NoStop}%
\bibitem [{\citenamefont {Bera}\ \emph {et~al.}(2014)\citenamefont {Bera},
  \citenamefont {Florens}, \citenamefont {Baranger}, \citenamefont {Roch},
  \citenamefont {Nazir},\ and\ \citenamefont {Chin}}]{bera_stabilizing_2014}%
  \BibitemOpen
  \bibfield  {author} {\bibinfo {author} {\bibfnamefont {S.}~\bibnamefont
  {Bera}}, \bibinfo {author} {\bibfnamefont {S.}~\bibnamefont {Florens}},
  \bibinfo {author} {\bibfnamefont {H.~U.}\ \bibnamefont {Baranger}}, \bibinfo
  {author} {\bibfnamefont {N.}~\bibnamefont {Roch}}, \bibinfo {author}
  {\bibfnamefont {A.}~\bibnamefont {Nazir}},\ and\ \bibinfo {author}
  {\bibfnamefont {A.~W.}\ \bibnamefont {Chin}},\ }\href@noop {} {\bibfield
  {journal} {\bibinfo  {journal} {Physical Review B}\ }\textbf {\bibinfo
  {volume} {89}},\ \bibinfo {pages} {121108(R)} (\bibinfo {year}
  {2014})}\BibitemShut {NoStop}%
\bibitem [{\citenamefont {Zaikin}\ and\ \citenamefont
  {Panyukov}(1986)}]{Zaikin}%
  \BibitemOpen
  \bibfield  {author} {\bibinfo {author} {\bibfnamefont {A.~D.}\ \bibnamefont
  {Zaikin}}\ and\ \bibinfo {author} {\bibfnamefont {S.~V.}\ \bibnamefont
  {Panyukov}},\ }\href@noop {} {\bibfield  {journal} {\bibinfo  {journal} {JETP
  Lett.}\ }\textbf {\bibinfo {volume} {43}},\ \bibinfo {pages} {670} (\bibinfo
  {year} {1986})}\BibitemShut {NoStop}%
\bibitem [{\citenamefont {Giamarchi}(2003)}]{Giam}%
  \BibitemOpen
  \bibfield  {author} {\bibinfo {author} {\bibfnamefont {T.}~\bibnamefont
  {Giamarchi}},\ }\href@noop {} {\emph {\bibinfo {title} {Quantum physics in
  one dimension}}}\ (\bibinfo  {publisher} {Oxford},\ \bibinfo {year}
  {2003})\BibitemShut {NoStop}%
\bibitem [{\citenamefont {Mizel}\ and\ \citenamefont
  {Yanay}(2020)}]{mizel_right-sizing_2019}%
  \BibitemOpen
  \bibfield  {author} {\bibinfo {author} {\bibfnamefont {A.}~\bibnamefont
  {Mizel}}\ and\ \bibinfo {author} {\bibfnamefont {Y.}~\bibnamefont {Yanay}},\
  }\href {https://doi.org/10.1103/PhysRevB.102.014512} {\bibfield  {journal}
  {\bibinfo  {journal} {Phys. Rev. B}\ }\textbf {\bibinfo {volume} {102}},\
  \bibinfo {pages} {014512} (\bibinfo {year} {2020})}\BibitemShut {NoStop}%
\bibitem [{\citenamefont {Bender}\ \emph {et~al.}(2020)\citenamefont {Bender},
  \citenamefont {Emonts}, \citenamefont {Zohar},\ and\ \citenamefont
  {Cirac}}]{Bender}%
  \BibitemOpen
  \bibfield  {author} {\bibinfo {author} {\bibfnamefont {J.}~\bibnamefont
  {Bender}}, \bibinfo {author} {\bibfnamefont {P.}~\bibnamefont {Emonts}},
  \bibinfo {author} {\bibfnamefont {E.}~\bibnamefont {Zohar}},\ and\ \bibinfo
  {author} {\bibfnamefont {J.~I.}\ \bibnamefont {Cirac}},\ }\href
  {https://doi.org/10.1103/PhysRevResearch.2.043145} {\bibfield  {journal}
  {\bibinfo  {journal} {Phys. Rev. Research}\ }\textbf {\bibinfo {volume}
  {2}},\ \bibinfo {pages} {043145} (\bibinfo {year} {2020})}\BibitemShut
  {NoStop}%
\bibitem [{\citenamefont {Bell}\ \emph {et~al.}(2016)\citenamefont {Bell},
  \citenamefont {Zhang}, \citenamefont {Ioffe},\ and\ \citenamefont
  {Gershenson}}]{bell_spectroscopic_2016}%
  \BibitemOpen
  \bibfield  {author} {\bibinfo {author} {\bibfnamefont {M.~T.}\ \bibnamefont
  {Bell}}, \bibinfo {author} {\bibfnamefont {W.}~\bibnamefont {Zhang}},
  \bibinfo {author} {\bibfnamefont {L.~B.}\ \bibnamefont {Ioffe}},\ and\
  \bibinfo {author} {\bibfnamefont {M.~E.}\ \bibnamefont {Gershenson}},\ }\href
  {https://doi.org/10.1103/PhysRevLett.116.107002} {\bibfield  {journal}
  {\bibinfo  {journal} {Phys. Rev. Lett.}\ }\textbf {\bibinfo {volume} {116}},\
  \bibinfo {pages} {107002} (\bibinfo {year} {2016})}\BibitemShut {NoStop}%
\bibitem [{\citenamefont {Murani}\ \emph {et~al.}(2020)\citenamefont {Murani},
  \citenamefont {Bourlet}, \citenamefont {le~Sueur}, \citenamefont {Portier},
  \citenamefont {Altimiras}, \citenamefont {Esteve}, \citenamefont {Grabert},
  \citenamefont {Stockburger}, \citenamefont {Ankerhold},\ and\ \citenamefont
  {Joyez}}]{Joyez}%
  \BibitemOpen
  \bibfield  {author} {\bibinfo {author} {\bibfnamefont {A.}~\bibnamefont
  {Murani}}, \bibinfo {author} {\bibfnamefont {N.}~\bibnamefont {Bourlet}},
  \bibinfo {author} {\bibfnamefont {H.}~\bibnamefont {le~Sueur}}, \bibinfo
  {author} {\bibfnamefont {F.}~\bibnamefont {Portier}}, \bibinfo {author}
  {\bibfnamefont {C.}~\bibnamefont {Altimiras}}, \bibinfo {author}
  {\bibfnamefont {D.}~\bibnamefont {Esteve}}, \bibinfo {author} {\bibfnamefont
  {H.}~\bibnamefont {Grabert}}, \bibinfo {author} {\bibfnamefont
  {J.}~\bibnamefont {Stockburger}}, \bibinfo {author} {\bibfnamefont
  {J.}~\bibnamefont {Ankerhold}},\ and\ \bibinfo {author} {\bibfnamefont
  {P.}~\bibnamefont {Joyez}},\ }\href
  {https://doi.org/10.1103/PhysRevX.10.021003} {\bibfield  {journal} {\bibinfo
  {journal} {Phys. Rev. X}\ }\textbf {\bibinfo {volume} {10}},\ \bibinfo
  {pages} {021003} (\bibinfo {year} {2020})}\BibitemShut {NoStop}%
\end{thebibliography}
\end{document}